\documentclass{article} 

\usepackage{url}

\newif\ifamsart
\amsartfalse 
\newif\ifieee
\ieeefalse 
\newif\ifllncs
\llncsfalse 

\newif\ifanonymized
\anonymizedfalse 

\newif\ifnosecimpl
\nosecimplfalse 

\ifllncs\usepackage{smallsubsub}\else
\fi

\usepackage{latexsym}
\usepackage{epic}
\usepackage{epsfig}
\usepackage{amsmath}
\usepackage{amssymb}
\usepackage{amscd}
\usepackage{url}
\usepackage{color}
\usepackage{wrapfig} 

\usepackage{graphics}
\usepackage[
hidelinks]{hyperref}

\definecolor{gray}{rgb}{.6,.6,.6}
\definecolor{darkgreen}{rgb}{.2,.6,.2}

\input{xy}
\xyoption{all}

\ifieee

\renewcommand\paragraph[1]{\par\bigskip\noindent\textbf{#1}\hspace{1ex}}
\newcommand\subparagraph[1]{\par\smallskip\noindent\textit{#1}\hspace{1ex}}

\fi

\newcounter{running}[section]
\newenvironment{renumerate}{\begin{enumerate}\setcounter{enumi}{\value{running}}}%
{\setcounter{running}{\value{enumi}}\end{enumerate}}

\newcommand{\cpsa}{\textsc{cpsa}}

\newcommand{\caif}{\textsc{caif}}
\newcommand{\fpga}{\textsc{fpga}}
\newcommand{\scmac}{\textsc{mac}}

\newcommand{\Adv}[1]{\mathsf{Adv}_{\textit{#1}}}

\newcommand{\kind}[1]{\ensuremath{\mathsf{#1}}}
\newcommand{\tagname}[1]{\ensuremath{\mathit{#1}\;}}

\newcommand{\ainstr}[1]{\ensuremath{\mathsf{#1}}}
\newcommand{\ainstrlog}{\ainstr{atlog}}
\newcommand{\ainstrescrstore}{\ainstr{protstore}}

\newcommand{\ainstrlogval}{\ainstr{iattest}}
\newcommand{\ainstrcheckval}{\ainstr{icheck}}
\newcommand{\ainstrescrowfor}{\ainstr{iprotect}}
\newcommand{\ainstrloadfrom}{\ainstr{iretrieve}}

\newcommand{\iattest}{\ainstrlogval}
\newcommand{\icheck}{\ainstrcheckval}
\newcommand{\iprotect}{\ainstrescrowfor}
\newcommand{\iretrieve}{\ainstrloadfrom}

\newcommand{\protfor}{\texttt{protfor}}
\newcommand{\retrfrom}{\texttt{retrvfm}}
\newcommand{\atloc}{\texttt{attestloc}}
\newcommand{\ckat}{\texttt{ckattest}}

\newcommand{\aattest}{\iattest}
\newcommand{\acheck}{\icheck}
\newcommand{\aprotect}{\iprotect}
\newcommand{\aretrieve}{\iretrieve}

\newcommand{\query}{\mathsf{query}}

\newcommand{\histfrom}{\twoheadleftarrow}    

\newcommand{\IS}{\ensuremath{\mathit{IS}}}
\newcommand{\CH}{\ensuremath{\mathit{sh}}}
\newcommand{\IF}{\textsc{if}}   

\newcommand{\src}{\ensuremath{\mathit{src}}} 
\newcommand{\dst}{\ensuremath{\mathit{dst}}} 
\newcommand{\svc}{\ensuremath{\mathit{svc}}}

\newcommand{\true}{\ensuremath{\mathsf{true}}}

\newcommand{\regular}{\ensuremath{\mathtt{compliant}}}

\newcommand{\dom}{\kind{dom}}
\newcommand{\ran}{\kind{ran}}

\newcommand{\qquote}[1]{\mbox{\texttt{"#1"}}}

\newcommand{\fn}[1]{\ensuremath{\operatorname{\mathit{#1}}}}

\ifamsart{}\else{

}\fi

\newcommand{\qdot}{\,\mathbf{.}\;}

\newcommand{\cons}{\mathbin{::}}
\newcommand{\seq}[1]{\langle #1 \rangle}

\newcommand{\dk}{\ensuremath{\kind{dk}}}
\newcommand{\dvk}{\ensuremath{\kind{dvk}}}
\newcommand{\sk}{\ensuremath{\kind{sk}}}
\newcommand{\vk}{\ensuremath{\kind{vk}}}

\newcommand{\shash}{\ensuremath{\mathit{sh}}}
\newcommand{\dhash}{\ensuremath{\mathit{dh}}}
\newcommand{\delhash}{\ensuremath{\mathit{dsh}}}
\newcommand{\setupdelhash}{\ensuremath{\mathit{suh}}}
\newcommand{\id}{\ensuremath{\mathit{imid}}}
\newcommand{\da}{\textit{DA}}
\newcommand{\oh}{\textit{arh}}
\newcommand{\kar}{k_{\textit{ar}}}

\newcommand{\valg}{\ensuremath{\mathop{svAlg}}}
\newcommand{\vh}{\ensuremath{\mathop{svh}}}
\newcommand{\vki}{\ensuremath{\mathop{vk}}}
\newcommand{\ski}{\ensuremath{\mathop{sk}}}

\newcommand{\anch}{\ensuremath{\mathit{anch}}}
\newcommand{\dstrh}{\ensuremath{\mathit{skdh}}}
\newcommand{\acth}{\ensuremath{\mathit{tgth}}}

\newcommand{\serial}{\ensuremath{\mathit{serial}}}
\newcommand{\ca}{\ensuremath{\fn{CA}}}

\newcommand{\trustchain}{\ensuremath{\mathit{trch}}}
\newcommand{\kdf}{\ensuremath{\kind{kdf}_h}}
\newcommand{\mac}{\ensuremath{\kind{mac}_h}}
\newcommand{\henc}{\ensuremath{\kind{enc}_h}}
\newcommand{\hdec}{\ensuremath{\kind{dec}_h}}

\newcommand{\payload}{\ensuremath{\mathit{payld}}}

\newcommand{\length}[1]{{| #1 |}}

\newcommand{\tagged}[2]{[\![\,#1\,]\!]_{#2}}

\newcommand{\hash}[1]{\#({#1})}

\newcommand{\tee}{\textsc{tee}}

\newcommand{\bluenode}{\textcolor{blue}{\bullet}}

\newcommand{\blacknode}{\textcolor{black}{\bullet}}
\newcommand{\graynode}{\textcolor{gray}{\bullet}}

\ifllncs
\newtheorem{prop}{Proposition}{\bfseries}{\itshape}
\newtheorem{principle}{Principle}{\bfseries}{\itshape}
\newtheorem{mylemma}[prop]{Lemma}{\bfseries}{\itshape}
\newtheorem{thm}{Theorem}{\bfseries}{\itshape}
\newtheorem{myrule}{Rule}{\bfseries}{\upshape}
\newtheorem{eg}{Example}{\bfseries}{\upshape}
\newtheorem{cor}[prop]{Corollary}{\bfseries}{\itshape}
\newtheorem{corsep}{Corollary}{\bfseries}{\itshape}

\spnewtheorem*{thmredux}{Theorem}{\bfseries}{\itshape}
\spnewtheorem*{propredux}{Proposition}{\bfseries}{\itshape}

\else
\ifieee\makeatletter
\def\@begintheorem#1#2{\@IEEEtmpitemindent\itemindent\relax\topsep 0pt\rmfamily\trivlist%
    \item[]\noindent\textbf{#1\ #2:} \itemindent\@IEEEtmpitemindent\relax}
\def\@opargbegintheorem#1#2#3{\@IEEEtmpitemindent\itemindent\relax\topsep 0pt\rmfamily \trivlist%
    \item[]\textit{\indent #1\ #2\ (#3):} \itemindent\@IEEEtmpitemindent\relax}
\makeatother
\fi
\newtheorem{theorem}{Theorem}

\newtheorem{definition}{Definition}

\newtheorem{cor}{Corollary}
\newtheorem{corsep}{Corollary}
\newtheorem{lemma}{Lemma}

\ifamsart{}\else
\def\squareforqed{\hbox{\rlap{$\sqcap$}$\sqcup$}}
\def\qed{\ifmmode\squareforqed\else{\unskip\nobreak\hfil
\penalty50\hskip1em\null\nobreak\hfil\squareforqed
\parfillskip=0pt\finalhyphendemerits=0\endgraf}\fi}

\newenvironment{proof}{\par\smallskip{\noindent\rm\bfseries
    Proof:}}{
  \par
}
\fi 
\fi

\def\defsymbol{\ensuremath{\scriptstyle /// }}
\def\defend{\ifmmode\squareforqed\else{\unskip\nobreak\hfil
\penalty50\hskip1em\null\nobreak\hfil\defsymbol
\parfillskip=0pt\finalhyphendemerits=0\endgraf}\fi}

\newenvironment{proofsketch}{\par\smallskip{\noindent\rm\bfseries
    Proof sketch:}}{\par\smallskip\noindent}

\ifieee

\fi

\usepackage{fancyhdr}

\fancypagestyle{prstitlepage}{%
  \fancyhf{}%
  \fancyhead[L]{\centering\textbf{Approved for Public Release;
      Distribution Unlimited. Public Release Case Number 24-2377}}
  \fancyhead[R]{}%
  \fancyfoot[L]{\centering \bf NOTICE

  Portions of this technical data were produced for the
  U. S. Government under Contract No. FA8702-19-C-0001 and
  W56KGU-18-D-0004, and is subject to the Rights in Technical
  Data-Noncommercial Items Clause DFARS 252.227-7013 (FEB 2014)

  \copyright 2024 The MITRE Corporation}%
}

\fancypagestyle{citationtitlepage}{%
  \fancyhf{}%
  \fancyhead[L]{\centering{Appearing in:  2025 IEEE Symposium on
      Computer Security Foundations}}%
  \fancyhead[R]{}%
  \fancyfoot[L]{\centering{This version in single column format for
      readability;\\
      table of contents on p.~\pageref{toc:page}.}} %
}

\fancypagestyle{prs}{%
  \fancyhf{}%
  \fancyhead[L]{PRS case 24-2377}
  \fancyhead[R]{\thepage}%
  \fancyfoot[L]{Copyright {\copyright} 2024, The MITRE Corporation.
    All rights reserved.}%
}

\title{Cryptographically Assured Information Flow:  Assured Remote
  Execution\thanks{Copyright {\copyright} 2024, The MITRE Corporation.
    All rights reserved.}
}
\author{Scott L. Dyer, Christian A. Femrite, \\ Joshua
  D. Guttman,\thanks{Corresponding author, \texttt{guttman@mitre.org}.
    This work was funded by MITRE's Independent Research and
    Development Program.}  ~ Julian P. Lanson, Moses D. Liskov \\[2mm]
  The MITRE Corporation
}

\ifllncs
  \institute{The MITRE Corporation}
\fi

\begin{document}
\maketitle
\thispagestyle{citationtitlepage}
%
%
%

\begin{abstract}
  \emph{Assured Remote Execution} for a device is the ability of
  suitably authorized parties to construct secure channels to
  processes executing known code running on that device.  Assured
  remote execution requires hardware including cryptographic
  primitives.

  \emph{Cryptographically Assured Information Flow} ({\caif}) is a
  hardware mechanism that enables Assured Remote Execution.  It is
  akin to existing Trusted Execution Environments, but securely
  implements an ideal functionality for \emph{logging} and
  \emph{confidential escrow}.  Symbolic protocol analysis demonstrates
  that Assured Remote Execution is achieved even against a strong
  adversary able to modify our programs and execute unauthorized
  programs on the device.

  Assured remote execution allows trustworthy remote attestation, and
  a core part of secure remote reprogramming.\ifanonymized{}\else{
      \ifieee{\footnote{{Copyright {\copyright} 2024, The MITRE
              Corporation.
              All rights reserved.}  \\
            Corresponding author:  Joshua Guttman,
            \texttt{guttman@mitre.org}.}}  \fi} \fi

%
%
%

\end{abstract}


\section{Introduction}
\label{sec:intro}

Suppose you have control of a device $d$ early in its life, after
which $d$ may be physically inaccessible, e.g.~on a satellite, or
rarely accessible, e.g.~one of many devices on ships, or embedded in
airplanes, or scattered throughout the electric power grid.
Long-term, can you deliver messages exclusively to specific, known
processes executing on $d$?  Can you, when receiving a message, be
sure it was prepared by a specific, known process on $d$?  Can the
processes run code written and delivered long after $d$ was
initialized?

This is the \emph{Assured Remote Execution} challenge.

Assured remote execution requires hardware support, 
as well as cryptography to protect messages in transit and to ensure
authenticity of the endpoint $d$ and active process within $d$.
%
%
Thus, solving the assured remote execution challenge requires both
device-local mechanisms on $d$ and distributed mechanisms to
coordinate $d$ with its owner or peers.  This paper offers a
device-local mechanism in
{\S\S\,}\ref{sec:ideal}--\ref{sec:properties}, and shows that it
suffices for protocols to coordinate with $d$ starting in
{\S\,}\ref{sec:strategy}.

Good solutions should:  \nocite{FarmerGuttmanSwarup96}
\begin{enumerate}
  \item \label{item:good:solutions:hw} Use a simple hardware
  basis \label{item:good:solutions:crypto} relying only on simple,
  efficient, well-understood crypto primitives;
  \item \label{item:good:solutions:adversary} Achieve the assured
  remote execution even against a strong adversary capable of running
  its own software on the device, or modifying existing software,
  including hypervisor software and software running during boot;
  \item \label{item:good:solutions:verification} Yield a verification
  strategy for using the mechanism, including the assured remote
  execution protocols.
\end{enumerate}

We define here a hardware basis adapted from existing Trusted
Execution Environments.  It uses cryptography to satisfy an ideal
functionality controlling information flow among processes local to
$d$, where we identify processes by the hash of their executable code
and constants.  This allows a process on $d$ receiving certain data
values to identify the processes that generated them, and it allows a
process on $d$ wanting to pass a data value confidentially to a
particular recipient process to do so without any other process
observing the value.  We call it {\caif}, for \emph{Cryptographically
  Assured Information Flow}.

Our hardware basis uses only hashing, key derivation functions,
message authentication codes ({\scmac}s), and authenticated symmetric
encryption.  These form a small collection of deeply understood
primitives, meeting criterion~\ref{item:good:solutions:hw}.

We have designed our mechanisms under the assumption that some
processes on our devices may run carefully vetted, trustworthy code,
whereas others may run questionable or even malicious code.  We do not
assume protected storage to hold executables within a {\caif} device,
so our conclusions hold even if an adversary modifies our programs in
the filesystem or installs their own.  Hence, higher level software
can use {\caif} to help prevent malicious execution without circular
dependencies.  So our adversary model meets
criterion~\ref{item:good:solutions:adversary}.

An ideal functionality~\cite{GGM86,canetti2001universally}
characterizes {\caif}, and our cryptographic mechanism
\emph{simulates} it to within negligible probability
(Cor.~\ref{cor:caif:if:indistinguishable}).  Lemmas saying the ideal
functionality enforces our intended information flow constraints are
thus also near approximations to the cryptographic mechanism.

Having justified the {\caif} mechanism, we show how to build assured
remote execution on top of it.
We formalize the behaviors as protocols in {\cpsa}, a symbolic-style
Cryptographic Protocol Shapes Analyzer that supports both message
passing and local device state~\cite{cpsa23}.  {\cpsa} helped us
eliminate errors, discover core ideas, and assure that the resulting
mechanisms satisfy our security claims.  Our {\cpsa} models
incorporate a strong adversary that can run any code, subject to the
assumption that code that yields the same hash value under a strong
hash function will also yield the same computational behavior.  The
ideal functionality proof and symbolic protocol analysis together meet
criterion~\ref{item:good:solutions:verification}.

\paragraph{The core idea of CAIF.}  {\caif} provides two central
functions.  One enables a \emph{service} (certain processes) to
\emph{log} itself as the source or authorizer of a data item.  Other
parties can subsequently \emph{check} whether an expected service has
logged a data item.  The other function enables a service to
\emph{escrow} a piece of data for a service as recipient.  Only that
recipient can then \emph{retrieve} it.
In a \emph{check} or \emph{retrieve} operation, the recipient of a
logged or escrowed value names the value's expected source.  The
operation fails if that source did not \emph{log} or \emph{escrow} it,
so success guarantees the value's provenance.
%

The cryptography-free {ideal functionality} of {\S\,}\ref{sec:ideal}
implements these functions via an \emph{unbounded secure memory},
proving desirable behavioral properties
(Lemmas~\ref{lemma:log:behavior}--\ref{lemma:escrow:secrecy}).  This
secure memory holds the logged associations of service and data, and
the escrowed associations of data, source, and recipient.

{\caif} devices ({\S\,}\ref{sec:caif}) use {\scmac}s for logging and
authenticated encryption for data escrow.  They require only one
unshared ``intrinsic secret'' $\IS$, used as an input to key
derivation for those cryptoprimitives.  {\caif} devices name services
by the hash of their executable code, ensuring that two services with
the same name will have the same computational behavior.

\paragraph{Assured remote execution} requires evidence a service
$\svc$ on $d$ sent messages $m$.  Our scheme uses a signing keypair
$(\ski,\vki)$ with a certificate chain for the verification key
$\vki$.  The certificate chain provides public evidence that $\ski$ is
escrowed for $\svc$, and that the provenance of $\ski$ leads back to a
previous service $\svc_0$ that escrows it \emph{only} for $\svc$.  At
the root of the chain, a certifying authority authenticated $d$ using
a shared key that was established early in $d$'s life.

If $m$ is signed with $\ski$, then $\svc$ bears responsibility
(see~\ref{sec:delegation:goal}).  For confidential channels to $\svc$,
the signed messages $m$ can be used for key
encapsulation~\cite{shoup2001proposal}.

\paragraph{Contributions.}  We make three main contributions.
\begin{enumerate}
  \item We define {\caif} and its ideal functionality for data logging
  and escrow ({\S\S\,}\ref{sec:ideal}, \ref{sec:caif}).
  \item We prove that {\caif}, if using strong cryptography, is
  computationally indistinguishable from an instance of the ideal
  functionality ({\S\,}\ref{sec:properties}).
  \item We develop a sequence of protocols on {\caif} to achieve
  assured remote execution.
  %
  %
  Symbolic protocol analysis shows they achieve this goal despite a
  strong adversary that can execute code of its own choice on our
  devices.
  %
  
  {\S\,}\ref{sec:strategy} gives our strategy, and
  {\S\S\,}\ref{sec:anchoring}--\ref{sec:delegation} provide
  details.\label{item:contrib:protocols}

\end{enumerate}
%
%
%
{\caif}'s guarantees are independent of delicate systems-level
considerations, such as how software obtains control at boot.
{\S\,}\ref{sec:challenges} identifies key challenges and a preliminary
use case that is simpler than assured remote execution, which
{\S\,}\ref{sec:using} shows how to meet.
{\S\S\,}\ref{sec:related}--\ref{sec:conc} discuss related work and
conclude.




\section{Current challenges}
\label{sec:motivation}
\label{sec:challenges}

%
%

{\caif} is motivated by several ingredients in the current situation
for cryptographic devices and secure systems design.  

\subsection{Background challenges}
\label{sec:motivation:challenges}

\paragraph{The quantum-resistant transition.}  Motivated by the
quantum cryptanalytic threat, new quantum-resistant primitives are now
in draft standard~\cite{NIST2022,FIPS203,FIPS204,FIPS205}.
However, {\caif}'s guarantees are independent of asymmetric
cryptography such as digital signatures.  Long-lived {\caif} devices
meet their guarantees even if these primitives are broken and revised,
or if their key sizes must be adjusted.  New asymmetric algorithms and
root-of-trust keys---typically, signature verification keys---can be
installed securely on geographically dispersed {\caif} devices,
yielding a long-term security architecture depending only on stable,
efficient symmetric cryptographic primitives.

\paragraph{Trusted Execution Environments.}  If asymmetric algorithms
may need to evolve, existing Trusted Execution Environments
({\tee}s)~\cite{cryptoeprint:2016:086,IntelTDX,amdSEV21} are not the
right tool.  Although they use only symmetric cryptography at the
hardware level, they rely on public key encryption to protect data
passing from one enclave to another.  This was acceptable when quantum
cryptanalysis seemed distant, but is no longer.

{\caif} contrasts in two central ways with prior {\tee}s.  First,
prior {\tee}s construct a device-local secret unique to each enclave,
and deliver this to the enclave for actions including local
attestations.  {\caif} constructs such a key but instead itself
generates {\scmac} tags to log data.  The {\tee} behavior does not
satisfy a logging ideal functionality, and does not allow us to
construct it:  Some enclaves may choose to disclose their key,
producing counterexamples to Lemma~\ref{lemma:log:behavior}
(cf.~{\S\,} \ref{sec:related:tees}).  Second, {\caif}'s escrow also
constructs keys for ordered \emph{pairs} of services, and uses those
for its confidential escrows.  This provides a symmetric method for
service-to-service information flow, which existing {\tee}s entirely
lack.

\subsection{A CAIF application:  Satellite reprogramming}
\label{sec:motivation:example}
\label{sec:challenges:example}

{\caif} pays off for widely dispersed, long-lived devices with clear
security goals and programs that may need to evolve.

For instance, the owner of a network of communications satellites
needs to manage them securely from the ground through decades of use.
The satellites must also create secure connections among themselves.
Customers and other network providers also need secure connections
with them.

These secure connections may need updated asymmetric cryptography
within the lifetime of the satellites as quantum resistant algorithms
or key sizes may evolve.  New crypto code with new root-of-trust keys
must then be installed while the satellites and their crypto hardware
are aloft.  Data integrity is critical:  An adversary whose bogus root
key is accepted can command the satellite.  Thus, services can use a
root key only if it must have been installed by the authorized local
service, which in turn authenticated its origin from the satellite's
management on the ground.  Local provenance leads back to a root key
installer, from which an authenticated protocol must lead back to the
management.

\paragraph{Updating signature algorithm and root-of-trust key.}
Suppose a {\caif}-equipped device $d$, after some initial terrestrial
preparation, called ``anchoring,'' is loaded into a satellite to
provide cryptographic functionality and lofted into orbit; the
operator of the satellite wants to exert control over the satellite
via $d$.  This operator is called the \emph{device authority} $\da$.
We assume for now that $\da$ shares a long-term secret $\kar$ which
can be used only by a particular service with code hash $\oh$;
{\S\,}\ref{sec:anchoring:distr} sets up $\kar$ via the terrestrial
preparation.

Years later, the signature algorithms $d$ uses may need updating;
maybe the old ones are already compromised.  $\da$ can send new code
$\valg$ for signature verification to $d$ via ordinary communications,
but must also send evidence the new code is trustworthy and received
unaltered, as well as a new root-of-trust signature verification key
$\vki$ for checking signatures with $\valg$.  In {\S\,}\ref{sec:using}
we show how to use {\caif} to do so securely.   This is a smaller
challenge than assured remote execution, which we will turn to in
{\S\S\,}\ref{sec:strategy}--\ref{sec:delegation}.  



\section{An Ideal Functionality for CAIF}
\label{sec:ideal}

We characterize {\caif} with an \emph{ideal functionality} {\IF},
meaning a well-defined set of behaviors that might be difficult to
achieve directly.  {\IF} would require an unbounded amount of memory
under its exclusive control, about which no observer gains any
information except through {\IF}'s official interface.
%
%
Lemmas~\ref{lemma:log:behavior}--\ref{lemma:escrow:secrecy} prove
desirable behavioral properties of the {\IF}.

In Section~\ref{sec:caif} we will introduce {\caif} devices using
cryptography, and in Section~\ref{sec:properties} we prove that these
{\caif} devices offer a near approximation to the {\IF}'s behavioral
properties.


\subsection{Main elements}
\label{sec:ideal:central}

We consider a system as a collection of \emph{active processes} that
act by executing instructions.  Some active processes are
distinguished as \emph{services}.  A service has an unchanging
executable code segment, and an unshared heap for private
computations.  Because the code segment is unchanging, its hash serves
as a persistent \emph{principal} or \emph{identity} for the service,
and also determines its computational behavior.

The instruction set includes two pairs of special instructions besides
normal computational steps.  The first pair allows a service to
\emph{log} itself in an attestation log $\ainstrlog$ as source or
authority for data, so other active processes can later make decisions
based on its provenance:
\begin{description}
  \item[\iattest] has one parameter, which points to a region of data
  with some contents $v$.  The logging functionality selects a
  \emph{tag}, a bitstring $\tau$, and stores a record into
  $\ainstrlog$ associating the currently active service identity $P_s$
  with $v$ and the tag $\tau$.
  %
  %
  Logging returns $\tau$ in response.
  \item[\icheck] has three parameters, namely a service principal
  identity $P_s$, a pointer to a region of data with some contents
  $v$, and a tag $\tau$.  The logging functionality returns
  \emph{true} if the named service $P_s$ previously logged $v$ into
  $\ainstrlog$ via {\iattest}, with tag $\tau$.  Otherwise, it returns
  \emph{false}.
\end{description}
Any active process can use {\icheck} to see if $P_s$ has logged itself
as an authority for $v$.  However, only a service can execute
{\iattest}, since only services have a persistent identity $P_s$.
Some function $F_{\mathit{log}}$ of $P_s$ and $v$ determines $\tau$.
To implement {\iattest}, one would use a {\scmac} as
$F_{\mathit{log}}$, so $\tau$ is the {\scmac} tag.  The tag $\tau$ and
$v$ may be passed from source to recipient through shared resources
such as a file-system.

The second pair of instructions ensures provenance of the source, and
also provides data \emph{escrow} through a table $\ainstrescrstore$,
meaning that the source service $P_s$ is making the data $v$ available
just to one recipient service $P_r$:
\begin{description}
  \item[{\iprotect}] has two parameters, the intended recipient
  service principal identity $P_r$ and a pointer to a region of data
  with some contents $v$.  When executed by a currently active service
  identity $P_s$, the escrow functionality selects a randomized
  \emph{handle} $\eta$.  It stores a record in the lookup table
  $\ainstrescrstore$, indexed by $(\eta,P_s,P_r)$, pointing to the
  value $v$.  We write this $(\eta,P_s,P_r)\mapsto v$.

  The logging functionality returns $\eta$ in response.
  \item[{\iretrieve}] has two parameters, the expected source service
  principal identity $P_s$ and a handle $\eta$.  When executed by a
  currently active service identity $P_r$, the escrow functionality
  looks up the index $(\eta,P_s,P_r)$ in the table $\ainstrescrstore$.
  If any entry $(\eta,P_s,P_r)\mapsto v$ is present in
  $\ainstrescrstore$, that $v$ is returned to $P_r$.  Otherwise, it
  fails.
\end{description}
%
%
Lemmas~\ref{lemma:escrow:behavior}--\ref{lemma:escrow:secrecy} below
depend on {\iprotect}'s choice of handles $\eta$.
We assume it samples $\eta$ from a distribution $D_{j, P_s, P_r}$
for data of length $j$ escrowed by principal $P_s$ for recipient $P_r$
where:
\begin{enumerate}
  \item[(i)] the randomized choice of $\eta$ is independent of which
  $v$ is presented, for all $v$ of a given length $j$;
  \item[(ii)] For different lengths $j \not= j'$, the supports
  $\kind{supp}(D_{j, P_s, P_r})$ and $\kind{supp}(D_{j', P_s', P_r'})$
  are disjoint; and 
  \item[(iii)] The {\IF} does not re-use any handle $\eta$; it checks
  the table entries and re-samples in case of collision.
\end{enumerate}

\subsection{Behavioral lemmas about {IF}}
\label{sec:ideal:lemmas}

A strength of the ideal functionality definition is that several
properties of the \emph{behaviors} of an {\IF} follow easily from it.
A \emph{command} $c$ is an instruction together with a choice of its
command arguments, $v$ for {\ainstrlogval}; $(P_s, v, \tau)$ for
{\ainstrcheckval}; etc.
\medskip
\begin{definition}
  An \emph{event} is a triple (command,principal,result) of: a
  command; the executing service principal that causes it (or $\bot$
  if the active process is not a service); and the result of executing
  the instruction.


  A \emph{behavior} of a state machine $M$ equipped with principal
  identities is a finite sequence $\seq{(c_i,P_i,r_i)}_{i<\ell}$ of
  events in which each command $c_i$ can cause the result $r_i$ when
  executed by principal $P_i$ in some state that can arise from the
  preceding events $\seq{(c_j,P_j,r_j)}_{j<i}$, starting from an
  initial state.

  An \emph{{\IF} behavior} is a behavior of {\IF} starting from the
  initial state with empty lookup tables.  \hfill ///
\end{definition}

\paragraph{Properties of IF:  Logging.}  We can summarize the
important properties of the attestation instructions in a lemma.  
It says that a check after a matching attest does yield true, and that
if a check yields true, then an earlier attest occurred.




\medskip
\begin{lemma}
  \label{lemma:log:behavior}
  Let $\alpha=\seq{(c_i,P_i,r_i)}_i$ be an {\IF} behavior.
  \begin{enumerate}
    \item If, for $i<j$, $c_i=\ainstrlogval(v)$ and
    $c_j={\ainstrcheckval}(P_i,v,r_i)$, then $r_j=\mathit{true}$.
    \item If $c_j={\ainstrcheckval}(p,v,\tau)$ and
    $r_j=\mathit{true}$, then for some $i<j$, $c_i=\ainstrlogval(v)$,
    $P_i=p$, and $r_i=\tau$.   \hfill /// 
  \end{enumerate}
\end{lemma}
%
%
This lemma
is independent of how $F_{\mathit{log}}$ chooses tags.
Lemma~\ref{lemma:log:behavior} makes no claim about sequential order;
\emph{check} confirms only presence not sequence relative to other
events.  Values $v$ containing hash chains can add sequential
information as usual.

\paragraph{Properties of IF:  Protection.}
The analog of Lemma~\ref{lemma:log:behavior} holds for
{\ainstrescrowfor}, using the re-sampling assumption (iii):

%
%
  %

\medskip 
\begin{lemma}
  \label{lemma:escrow:behavior}
  Let $\alpha=\seq{(c_i,P_i,r_i)}_i$ be an {\IF} behavior.
  \begin{enumerate}
    \item If, for $i<j$, $c_i=\ainstrescrowfor(P_j,v)$ and
    $c_j={\ainstrloadfrom}(P_i,r_i)$, then $r_j=v$.
    \item If $c_j={\ainstrloadfrom}(p,\eta)$ and $r_j=v$, then for
    some $i<j$, $c_i=\ainstrescrowfor(P_j,v)$, $P_i=p$, and
    $r_i=\eta$.  \hfill /// 
  \end{enumerate}
\end{lemma}

\paragraph{Ideal secrecy for {IF}.}
%
{\IF} leaks \emph{no information} about the values associated with
handles that are never retrieved.  
%
%
%

A \emph{schematic behavior} $\alpha_\nu$ in the variable $\nu$ results
from a behavior $\alpha$ by replacing one occurrence of a bitstring in
a command or result of $\alpha$ with the variable $\nu$.  If $b$ is
any bitstring, $\alpha_\nu[b/\nu]$ is the result of replacing the
occurrence of $\nu$ by $b$.  The latter may not be a behavior at all,
since this $b$ may be incompatible with other events in $\alpha$.

%
By (i), a strong, Shannon-style perfect secrecy claim holds for the
ideal functionality:
\begin{lemma}
  \label{lemma:escrow:secrecy}
  Let $\alpha_\nu=\seq{(c_i,P_i,r_i)}_i$ be a schematic behavior,
  where $\nu$ occurs in an {\ainstrescrowfor} instruction
  $c_i=\ainstrescrowfor(P_r,\nu)$ executed by $P_s$.  Let $\ell$ be a
  length of plaintexts for which the result $r_i$ is possible.  By
  assumption (ii), there is a unique such $\ell$.  Let $D$ be any
  distribution with $\kind{supp}(D)\subseteq \{0,1\}^\ell$.

  Suppose there is no subsequent $c_j={\ainstrloadfrom}(P_i,r_i)$ with
  this input $r_i$ and $P_j=P_r$.
  For every $b \in \kind{supp}(D)$:
  \begin{enumerate}
    \item $\alpha_\nu[b/\nu]$ is a behavior; 
    \item the probability
    $Pr[v_0\leftarrow D; v_0=b \mid \alpha_\nu[v_0/\nu]\, ]$ that the
    given $b$ was sampled from $D$ conditional on observing
    $\alpha_\nu[v_0/\nu]$ equals $Pr[v_0\leftarrow D; v_0=b ]$.
    \hfill ///
  \end{enumerate}
\end{lemma}
%

\medskip
Lemmas~\ref{lemma:log:behavior}--\ref{lemma:escrow:behavior} are
authentication properties; Lemma~\ref{lemma:escrow:secrecy} is a
secrecy property.  Lemma~\ref{lemma:escrow:secrecy} is strong; since
{\caif} must approximate it using concrete cryptography, it achieves
only a computational approximation to it.  The {\IF} is parameterized
by a function $F_{\mathit{log}}$ and a family of distributions
$D_{\ell, P_s, P_r}$; each instance of {\IF} is of the form
\IF$[F_{\mathit{log}}, \{D_{\ell, P_s, P_r}\}]$.  The lemmas hold for
all values of these parameters satisfying (i)--(iii).
%


\section{Using CAIF}
\label{sec:using}

\subsection{Satellite reprogramming via CAIF}
\label{sec:example:using:caif}

We turn back to our satellite reprogramming challenge for {\caif} from
{\S\,}\ref{sec:challenges:example}.
%
We use the assumed shared secret $\kar$ as a {\scmac} key to
authenticate messages from $\da$.  {\scmac} suffices because we need
no confidentiality here; integrity, authentication, and authorization
are the goals.  In {\S\,}\ref{sec:strategy} we will start handling the
larger challenge of full assured remote execution, which requires
protecting long-term secrets.  

\begin{table}\ifieee{}\else\small\fi
  \centering
  \caption{Reprogrammable signature verification}\ifieee{}\else\vspace{2mm}\fi
  \begin{tabular}{llll}
    Svc & Hash & Secret & Actions \\[1mm]
    Auth.~Recip & $\oh$ & $\kar$ & $\mathsf{recv}\;
                                   \vh,\vki,\mac\, (\kar,(\vh,\vki))$ \\
        &&& $\iattest\;\qquote{ver}\, (\vh,\vki)$\\
        &&& $\iattest\;\qquote{cli}\, \vh$ \\[1mm]
    Sig.~ver & $\vh$ & -- & $\icheck\;\qquote{ver}\, (\vh,\vki)$ from
                            $\oh$ \\
        &&& $\mathsf{recv}\; m,\sigma$ \\
        &&& if~~$\tagname{verify}(\sigma,m,\vki)$  \\
        &&& then~~$\iattest\;\qquote{yes}\, (m,\sigma)$ \\ 
        &&& else~~$\iattest\;\qquote{no}\, (m,\sigma)$ \\[1mm]
    Clients & any & -- & $\mathsf{recv}\; m,\sigma$ \\
        &&& $\icheck\;\qquote{cli}\, \vh$ from $\oh$ \\
        &&& $\mathsf{send}\; m,\sigma$ to $\vh$ \\
        &&& if~~$\icheck\;\qquote{yes}\, (m,\sigma)$ from $\vh$ \\
        &&& then~~$\mathsf{accept}$ \\
        &&& else if~~$\icheck\;\qquote{no}\, (m,\sigma)$ \\
        &&& then~~$\mathsf{reject}$
  \end{tabular}
  \label{tab:reprogramming:ver}
\end{table}

On $d$, an \emph{authorized recipient} has service identity or code
hash $\oh$, and a \emph{signature verifier} with code $\valg$ has code
hash $\vh=\mathop{hash}(\valg)$.
\begin{description}
  \item[Authorized recipient] with hash $\oh$:  it receives an
  incoming message containing a code hash $\vh$ and a signature
  verification key $\vki$ and {\scmac}-checks it using $\kar$.  On
  success, it {\iattest}s a \emph{verifier record} containing $\vki$
  and $\vh$ and also {\iattest}s a \emph{client record} containing
  $\vh$.
  The verifier record {authorizes} $\vh$ to use key $\vki$.  The
  client record {authorizes} clients to use $\vh$ to verify
  signatures.

  The service $\oh$ is trusted to make these authorization claims.  As
  $\oh$ is independent of the signing algorithm, it may be installed
  at device initialization and left permanently unchanged.
  \item[Signature verifier] with hash $\vh=\mathop{hash}(\valg)$:  it
  obtains a verifier record containing a verification key $\vki$ and
  its own hash $\vh$.  If the record {\icheck}s as logged by $\oh$,
  then $\vh$ awaits requests $(m,\sigma)$ from clients.  If the
  signature verification succeeds for $(m,\sigma)$ with key $\vki$,
  then the service {\iattest}s a confirmation record containing
  $\mathsf{yes}$, $m$, and $\sigma$.  On failure, it {\iattest}s an
  error record.

  The authority $\da$ compiles the constant $\oh$ into the code
  $\valg$.  Altering the constant alters the hash $\vh$, so a log
  record with provenance from $\vh$ ensures the right origin $\oh$ was
  checked.
  \item[Clients,] when receiving a purported signed message
  $(m,\sigma)$, obtain a client record containing a hash value $\vh$,
  using {\icheck} to ensure it was logged by $\oh$.  They then request
  $\vh$ to verify the signature.  When a confirmation record
  containing $\mathsf{yes}$, $m$, and $\sigma$ is received, and it
  {\icheck}s successfully for $\vh$, the signature is valid.
  The clients' code also embeds $\oh$ as a constant.
\end{description}
If $\da$ authorizes \emph{only} verifier code 
acting as described, then messages will be accepted only if validly
signed.
%
%
Symbolic protocol analysis confirms this, assuming $\kar$ is protected
from compromise at $\da$.

\ifnosecimpl{}\else{
    The symbolic protocol analysis makes some fine points explicit.
    For instance, $\da$ authorizes more than one root verification key
    $\vki$ for the code $\vh$, the client does not learn which key
    verified a particular $(m,\sigma)$.  Adding $\vki$ to the
    confirmation record achieves this, if desired.}  \fi

A remote authority can use this method to enable a {\caif} device
to use new signature verification algorithms coded, delivered, and
authorized long after {\caif} device became physically inaccessible.
An initial anchoring event made the shared secret $\kar$ available
only to $\oh$ (see {\S\,}\ref{sec:anchoring:distr}).

\ifnosecimpl{}\else{
    \emph{Deauthorizing} an old, no longer secure signature algorithm,
    by contrast, requires an irreversible change, lest an adversary
    roll $d$ back to it.  A monotonic counter, judiciously used, would
    likely suffice, but that remains as future work.
  }
\fi

\paragraph{Assured remote execution?}  This protocol does not offer
assured remote execution:  It does not deliver evidence to a remote
observer that any $\svc$ is active on $d$.  It offers local client
services information provenance for $\vki$ and $\vh$, namely evidence
they were obtained by $\oh$, and are thus authorized if $\da$
safeguarded $\kar$.  the scheme authenticates messages from $\da$ to
$d$ assuming $\da$ protects its signing key.

Assured remote execution offers a converse, i.e.~authenticating
messages from a service $\svc$ on $d$ to external peers.  $\da$ can
use shared keys like $\kar$ for this authentication.  \emph{Signature
  delegation} ({\S\,}\ref{sec:delegation}) subsequently extends this
$\svc$-to-$\da$ assured remote execution to allow any peer to
authenticate any service on $d$ using (possibly new) digital signature
algorithms.  The underlying message authentication allows $d$ to get
signing keys for these services certified.

Access to long-term secrets such as $\kar$ at the $\da$ should be rare
and stringently restricted.
    %
We use them only to authorize new asymmetric algorithms on the device 
or to certify keys.  

\paragraph{CAIF for assured confidentiality.}  Signature verification
keys require integrity; by duality signing keys require
confidentiality.  A signing key that will authenticate a service on
$d$ must rely on the signing key remaining confidential;
adversary-installed code must not hijack the key.  Thus, local flow
should carry a $\da$-certified private signing key only to the code
authorized to use it.  $\da$ must know that the private signing key
cannot reach any further code as recipient.  We use the {\caif} escrow
operation {\iprotect} to achieve this; see~{\S\,}\ref{sec:delegation}.

\paragraph{Another application:  critical infrastructure.}  Devices
controlling electric grids, water systems, etc.~have succumbed to
widespread infiltration, unsurprisingly as their software is very
hardware-specific.  As they are long-lived and geographically
dispersed, hands-on reprogramming is impractical.  Remote methods are
needed to guarantee new code controls them.
Preliminary work with Field Programmable Gate Arrays
suggests {\caif}'s hardware burden is modest, making it feasible for
these small, cost-constrained devices.

\subsection{Techniques for effective CAIF use}
\label{sec:motivation:opportunities}
\label{sec:using:techniques}

A few core ideas underlie
%
%
effective {\caif} use.  
%
%
\ifnosecimpl{}\else{
    \paragraph{Local chains of provenance.}  Multistep chains of
    provenance can use {\caif} logging or escrow.  Suppose $\svc_0$
    logs $m_0$ that says, ``I computed $v_0$ from $v_1$, which was
    logged by $\svc_1$.''  Local services can {\icheck} the $m_0$ log
    entry, ensuring $\svc_0$ logged it.  If the $\svc_0$ code does not
    log $m_0$ unless it {\icheck}s $v_1$ and computes $v_0$, then
    $\svc_1$ logged $v_1$.  If $v_1$ is a message $m_1$ of similar
    form, we can follow the chain backward.

    The inference requires knowing what $\svc_0$'s code may do.
    %
    {\caif} does not help in determining this, but given evidence of
    it, {\caif} yields runtime conclusions about $v_0$ and $v_1$'s
    provenance.

    Confidential intermediate values delivered exclusively to $\svc_1$
    via escrow may also be chained together
    ({\S\,}\ref{sec:anchoring:trust:chains}).  The provenance
    information assures local services that the flow occurred, with
    \emph{some} confidential value $v_1$ escrowed with a handle
    $\eta_1$.

    %
    %
    \paragraph{How to use CAIF's local guarantees remotely.}
    Anchoring ({\S\,}{\ref{sec:anchoring}}) and signature delegation
    ({\S\,}{\ref{sec:delegation}}) yield authenticated messages from
    known services $\svc$ on $d$.  If $\svc$ reports locally validated
    provenance chains
    on $d$, the peer learns the source of $\svc$'s data, such as the
    pedigrees of secret keys.

%
    %
  }
\fi

\paragraph{But:  Hashes are unpredictable.}  When one service escrows
confidential data for a peer service, how does the source choose the
hash of its target service?  And how does a service decide which hash
stands for an authorized source service from which it should accept
data?  Hashes are fragile and do not reflect semantics.
There are a number of approaches:
\begin{enumerate}
  \item The hash $h_p$ of the intended peer service is embedded in the
  active service's executable code.  Alterations to $h_p$ in the code
  change its code hash $h_a$ to some other $h_a'$, rendering data such
  as keys escrowed for $h_a$ unavailable.\label{item:hashes:pre}

  If the peer is compiled first, the constant $h_p$ can appear in
  $h_a$'s executable.  Alternatively, several services may be
  constructed jointly with each others' hashes~\cite{ChenZ22}.
%
  \item The hash $h_p$ may be found within a logged or escrowed record
  $r$ the service reads, where {\caif} logged $h_0$ as the origin of
  $r$.  If $h_0$ is known to be trustworthy for a given purpose,
  e.g.~by method~(\ref{item:hashes:pre}), then trust for related
  purposes may be delegated to $h_p$ by $r$.\label{item:hashes:read}
  \item The hash $h_p$ may be received in a message authenticated as
  coming from an authority such as the device authority $\da$, who may
  delegate trust for a particular purpose to the service with hash
  $h_p$.\label{item:hashes:auth}

  The incoming message may be authenticated using a key $k$ found in
  an escrowed record with source hash $h_0$; $k$'s trustworthiness
  then comes via method~(\ref{item:hashes:read}).  Alternatively, an
  exchange occurring only in a protected environment can also
  authenticate the incoming message.
  \item The hash may be supplied as a parameter in an argument vector
  or in an unauthenticated incoming message.  A hash $h_p$ of bad code
  may be supplied.  This does no harm if $h_p$ is included in messages
  or attested records generated by the service, and the subsequent
  recipients can decide whether to trust
  $h_p$.\label{item:hashes:unauth}
  
  %
\end{enumerate}
In methods~(\ref{item:hashes:pre}) and~(\ref{item:hashes:auth}), the
authority
%
%
has had an opportunity to appraise whether the code with hash $h_p$
satisfies some behavioral property.  With
method~(\ref{item:hashes:unauth}), the recipient adapts its responses
to what it knows about that code.

We used method~(\ref{item:hashes:pre}) in
{\S\,}\ref{sec:example:using:caif}.
Method~(\ref{item:hashes:read}) appears in the client code of
{\S\,}\ref{sec:example:using:caif} and makes frequent appearances
subsequently.  We use method~(\ref{item:hashes:auth}) repeatedly in
{\S\S\,}\ref{sec:anchoring}--\ref{sec:delegation}, receiving hashes
$h$ in authenticated messages where the authentication depends on
escrowed keys.  Our anchoring protocol ({\S\,}\ref{sec:anchoring:ss})
uses method~(\ref{item:hashes:auth}) where the authentication depends
on a protected environment, which provides the fundamental trust basis
for our assured remote execution protocol.


Many combinations of the four methods are useful.

In our analyses below, we express behavioral properties as protocol
roles sending and receiving messages, and loading and storing records
into device state records.  These roles define what we are trusting
the programs to do, namely to act on data values of the relevant forms
only in the patterns codified in the roles.  Their behavior may vary
in other respects without undermining our conclusions.  Creating
trustworthy software is hard, but a precise description of what the
software will be trusted to do (and avoid doing) helps.

%
%


\paragraph{Some limits.}  The benefits of {\caif} apply only to
devices satisfying the specification in {\S\,}\ref{sec:caif}.  Hence,
manufacturers must prevent back doors; customers need reliable supply
chains from the manufacturer.
%
The customer must also anchor it properly (as in
{\S\,}\ref{sec:anchoring}), and protect the anchoring shared secret.

\ifnosecimpl{}\else{
    For different organizations' {\caif} devices to cooperate, each
    needs to know the peer organization's devices are
    {\caif}-compliant and were properly anchored.  Different network
    providers with {\caif}-equipped satellites can interoperate
    securely based on business agreements; interoperation among allies
    is also reasonable.  In a permissioned blockchain, the controlling
    authority can make contractual agreements with the parties.  Open
    access blockchains, by contrast, have no way to check which
    devices are in fact {\caif} devices.
  }
\fi

\paragraph{Information flow.}  The {\caif} mechanism allows services
to determine \emph{from which} service some data has come, and
\emph{to which} service it may be delivered.  In this sense, it
controls information flow.  However, it has different goals from
information flow in the sense of the large literature descending from
Goguen and Meseguer~\cite{GoguenMeseguer82}.  In particular, rather
than enforcing a system policy, our mechanism allows services to
enforce their own application-specific policies; it is not a mandatory
mechanism but a discretionary mechanism.  Moreover, unlike classical
noninterference, these policies are not ``transitive;'' service $s_2$
may accept data from $s_1$ and deliver it to $s_3$, whereas $s_1$ and
$s_3$ may refuse direct flow~\cite{rushby1992noninterference}.  Thus,
{\caif} enables discretionary, non-transitive information flow
control.

\paragraph{Terminology.}  
We write our hardware-based symmetric mechanisms as $\kdf(x)$,
$\mac(k,v)$, and $\henc(v,k)$ for the hardware key derivation
function, {\scmac}, and authenticated encryption.  The corresponding
hardware decryption is $\hdec(v,k)$.

We write $\tagged v k$ for a digital signature on message $v$ with
signing key $k$.  We will assume that $v$ is recoverable from
$\tagged v k$.  The latter could be a pair
$(v,\;\kind{dsig}(\kind{hash}(v),\,k))$ where $\kind{dsig}$ is a
digital signature algorithm.


A lookup table (e.g.~a hash table) is a set $T$ of index-to-result
mappings.  Each mapping takes the form
$\mathit{index}\mapsto\mathit{result}$.  $T$ satisfies the ``partial
function'' constraint:  if $i\mapsto r\in T$ and $i\mapsto r'\in T$,
then $r=r'$.  $T$'s domain and range are
$\dom(T)=\{i\colon\exists r\qdot i\mapsto r\in T\}$ and
$\ran(T)=\{r\colon\exists i\qdot i\mapsto r\in T\}$.

When $D$ is a distribution, we write $\kind{supp}(D)$ for its support,
i.e.~$\kind{supp}(D)=\{x\colon 0<\mathit{Pr}[y \leftarrow D; y=x] \}$.


\section{CAIF Devices}
\label{sec:caif}

We use cryptography to implement {\IF}, eliminating the protected
state in {\ainstrlog} and {\ainstrescrstore}.  This cryptographically
achieved {\IF} is {\caif}.  It uses a single fixed, unshared secret.

{\caif} is built around two main ingredients: first, the idea of a
{\caif} \emph{service}, a computational activity with a known
\emph{service hash} that serves as its identity
({\S\,}\ref{sec:caif:services}); and, second, two pairs of
\emph{instructions} or basic operations to ensure provenance and
control access of data passed among services
({\S\,}\ref{sec:caif:instructions}).  Auxiliary operations are also
needed to manage services ({\S\,}\ref{sec:caif:auxiliaries}).
{\S\,}\ref{sec:caif:devices} summarizes what being a {\caif} device
requires.

\subsection{CAIF control over services}
\label{sec:caif:services}

A {\caif} device designates some active processes as \emph{services}.
A service has an address space such that:
\begin{renumerate}
  \item Executable addresses are located only within a non-writable
  \emph{code segment};\label{item:if:svc:exec}
  \item A non-shared \emph{heap segment} is readable and writable by
  this service, but not by any other active
  process;\label{item:if:svc:heap}
  \item Other address space segments may be shared with other active
  processes.\label{item:if:svc:shared}
\end{renumerate}
These segments are disjoint, so that code is readable and executable
but not writable, while heap is readable and writable but not
executable.  A program can address them reliably, so that secrets
(e.g.) are written into unshared heap rather than shared memory.
Moreover:
\begin{renumerate}
  \item The {\caif} device controls when a service is active, and
  maintains the \emph{hash} of the contents of its \emph{code segment}
  as its \emph{service identity} or
  \emph{principal}.\label{item:if:svc:id}
\end{renumerate}
The code segment being immutable, the hash does not change, and
{\caif} regards it as a non-forgeable identity.  To refer to the
principal, we often speak simply of its \emph{service hash}.

%

\subsection{CAIF instructions}
\label{sec:caif:instructions}

{\caif} offers two pairs of instructions for service-to-service
information flow.  They correspond to {\ainstrlogval} and
{\ainstrcheckval} for asserting and checking provenance, and
{\ainstrescrowfor} and {\ainstrloadfrom} for escrowing data values and
controlling their propagation.  The primitive operations use symmetric
cryptography.  Their keys are derived from one or more service hashes
and a device-local unshared secret called the Intrinsic Secret.
\begin{renumerate}  
  \item An \emph{Intrinsic Secret} within each {\caif} device $d$ is
  shared with no other device or party, and is used exclusively to
  derive cryptographic keys for the {\caif} instructions.
\end{renumerate}
We write $\IS$ for this intrinsic secret.  $\IS$ is the only secret
that the {\caif} hardware has to maintain.  Hardware design can help
prevent $\IS$ being accessible except for key derivation.  It may be
implemented as a set of fused wires as in {\textsc{sgx}} or as a
Physically Unclonable Function as in
Sanctum~\cite{LebedevHoganDevadas18}.

The hardware furnishes four cryptographic primitives, namely a key
derivation function $\kdf$, a Message Authentication Code $\mac$, and
an authenticated symmetric encryption $\henc$ with the decryption
$\hdec$.  Thus, when $c$ was not generated by an invocation of
$\henc(p,k)$, then $\hdec(c,k)$ is negligibly likely to return a
non-failure result.

A conceptual view of the cryptographic hardware components is in
Fig.~\ref{fig:caif:crypto:hardware}.
\begin{figure}
  \centering{\tiny
    \[
      \xymatrix@R=1mm@C=4mm{
        &*+[F]{\;\txt{IS}\;}\ar@3[dd]&&&&&
        \\
        &&&&&& 
        \\
        &*+=[F]{\;\txt{
            KDF}\;}\ar@3[ddrr]\ar@3[dddddl]&&&&&{\iftrue c, \CH_s, \CH_r\else\null\fi}\ar[lllll]    \\
        &&&&&*+[F]{\;\txt{IV src}\;}\ar@3[dll]&    \\
        &&&*+=[F]{\;\txt{Sym \\ Enc-Dec}\;}\ar[dd]&&& {\iftrue v\mbox{ or }\henc({v},k)\else\null\fi}\ar[lll]   \\
        &&&{\iffalse\txt{\tiny data}\else\null\fi} &&&    \\
        &&&&&& 
        \\
        *+=[F]{\;\txt{{\scmac}}\;}\ar[dd]&&&&&& {\iftrue v\else\null\fi}\ar[llllll]   \\
        &&&&&&  \\
        {\iffalse\txt{\tiny data}\else\null\fi} &&& &&& }
    \]}
  \caption[Hardware Crypto]{{\caif} Hardware Cryptography Components}
  \label{fig:caif:crypto:hardware}
\end{figure}
Triple arrows are buses; their source and destination are guaranteed
by physical connections.

Instructions may \emph{fail} or \emph{succeed}.  Failing may be
implemented by a transfer of control or terminating the process, or
simply by setting a condition code to be checked to determine whether
to branch in subsequent instructions.

\paragraph{Provenance via MACs.}  One pair of primitive instructions
uses {\scmac}s for \emph{attesting locally} and \emph{checking an
  attestation}; they identify a service that has generated or endorsed
a data value $v$, typically the data in a region of unshared heap
defined by a pointer and a length.

%
\begin{description}
  \item[\normalfont\atloc$(v)$:]  Computes a {\scmac} on given data
  $v$ using a key derived from $\IS$ and the service hash $\CH$ of the
  current service.  If no service is currently executing it fails.
  
  The {\scmac} key $k$ is the result of key derivation via $\kdf$:
  \[ k = \kdf(\mbox{\texttt{"at"}}, \; \IS, \; \CH) .
  \]
  The instruction computes the {\scmac} $m=\mac(k,v)$ of $v$ with $k$
  and makes $m$ available.  The {\scmac} tag $m$ and $v$ may
  subsequently be copied anywhere.
  \item[\normalfont\ckat$(\CH_s,v,m)$:]  Checks---given a service hash
  $\CH_s$ of the purported source, data $v$, and a purported {\scmac}
  $m$---whether $m$ is correct.  It returns true or false depending
  whether the purported {\scmac} $m$ equals a recomputed {\scmac}.  Thus,
  letting:
  \[ k = \kdf(\mbox{\texttt{"at"}}, \; \IS, \; \CH_s)
  \]
  the result is true iff
  \( m=\mac(k,v) \).  
\end{description}
A service may thus log itself as the origin or approver of $v$;
%
%
any recipient of
$v$ and $m$, if executing on the same device with the same intrinsic
secret $\IS$, can subsequently ascertain its provenance.  More
specific ``intent'' may be encoded into the content of $v$, which may
then be copied to a shared resource, e.g. a filesystem.  Thus,
{\atloc} and {\ckat} provide a device-local mechanism for asserting
and confirming provenance.

\paragraph{Protection and provenance via encryption.}  The remaining
primitive operations \emph{protect} a value \emph{for} a named
recipient by authenticated encryption, and \emph{retrieve} a value
\emph{from} a named source by decryption.  The plaintext should always
be located within unshared heap.  The ciphertext can pass freely
through shared resources, but only the stipulated recipient recovers
its content, and only if it was from the expected source.
\begin{description}
  \item[\normalfont\protfor$({\CH}_r,v)$:]  Computes---given an
  intended recipient's service hash ${\CH}_r$ and a data value
  $v$---an authenticated symmetric encryption of $v$ using a key $k$
  derived from $\IS$, the service hash ${\CH}$ of the currently
  executing service, and ${\CH}_r$.  If no service is currently
  executing it fails.

  The encryption key $k$ is computed via $\kdf$ as:
  \[ k = \kdf(\mbox{\texttt{"pf"}}, \; \IS, \; \CH, \;{\CH}_r) .
  \]
  The third component is the service hash ${\CH}$ of the currently
  running service, and ${\CH}_r$ is its intended recipient.  The
  instruction computes the encrypted value:
  \[ e = \henc({v},{k})
  \]
  which is stored back into a suitable region of memory.  The
  resulting $e$ may subsequently be copied anywhere.  
  
  \item[\normalfont\retrfrom$({\CH}_s,e)$:]  Decrypts $e$---given an
  expected source's service hash ${\CH}_s$ and a purported encryption
  $e$---using key $k$ derived from $\IS$, the source's hash ${\CH}_s$,
  and the code hash ${\CH}$ of the current service.  Fails if the
  (authenticated) decryption fails, i.e.~if the associated tag is
  wrong, or if no service ${\CH}$ is executing.
 
  The decryption key $k$ is computed via $\kdf$ as:
  \[ k = \kdf(\mbox{\texttt{"pf"}}, \; \IS, \; \CH_s, \;{\CH}) .
  \]
  The service hash ${\CH}$ of the currently executing service is now
  the last component, and the source $\CH_s$ is the previous one.  If
  $v = \hdec({e},{k})$, the plaintext $v$ is stored back into unshared
  heap.
\end{description}
%

Local provenance with access control results from {\protfor} and
{\retrfrom}.  Only the service specified in the {\protfor} learns
anything from the result, and only on the same device.  If ${\CH}$ or
$\IS$ differs, the $k$ in {\retrfrom} will differ, causing the
authenticated decryption to fail.  If ${\CH}$ specifies the wrong
source ${\CH}_s$, $k$ will again differ and failure ensue.

%
%

%
%

\subsection{Auxiliary operations}
\label{sec:caif:auxiliaries}

We also need some auxiliary operations on services:
\begin{description}
  \item[\normalfont\texttt{create-service}:]  Creates a service with
  the code contained in a buffer of memory, plus some other resources
  including a newly allocated unshared heap.  The service hash is
  ascertained to be used by the {\caif} mechanism.
  
  The resulting service does not execute immediately, but is placed on
  a list of runnable services.
  \item[\normalfont \texttt{start-service}:]  Given a runnable
  service, start it executing with access to the values of any
  additional parameters.
  \item[\normalfont \texttt{yield}:]  Stop executing the current
  service, retaining its state for future \texttt{start-service}s.
  \item[\normalfont \texttt{exit-service}:]  Zero the unshared heap of
  the service and eliminate it from the list of runnable services.
\end{description}
Any binary executable may become a service, with or without {\caif}
instructions; that binary executable is then indelibly associated with
it through the service hash.  Thus, its identity is correctly
reflected however it uses the four core {\caif} instructions.
Varying implementations can use different versions of the auxiliary
instructions.

\subsection{CAIF devices}
\label{sec:caif:devices}

By a \emph{CAIF device} we mean a hardware device that enables the
{\caif} instructions in {\S\,}\ref{sec:caif:instructions} to be
performed by services of the kind given in
{\S\,}\ref{sec:caif:services}; the auxiliary operations in
{\S\,}\ref{sec:caif:auxiliaries} may be carried out using a
combination of software and hardware.



\begin{definition}
  A {\caif} \emph{state} is an intrinsic secret $\IS$.  No {\caif}
  command modifies the {\caif} state.

  {\caif} \emph{behaviors} $\seq{(c_j,p_j,r_j)}_{j<i}$ are behaviors
  containing {\caif} commands $c_j$, active principals---i.e.~code
  hashes---$p_j$ and command results $r_j$.
\end{definition}

\paragraph{Cryptographic choices.}  {\caif} devices should be equipped
with strong cryptographic primitives.  In particular, implementors
should endeavor to ensure:
\begin{description}
  \item[\normalfont Key derivation:]  The key derivation function
  $\kdf$ is indistinguishable from a random function.
  We measure it by the \emph{pseudorandomness advantage}
  $\Adv{prf}(A,k,q)$ of any adversary algorithm $A$ aiming to separate
  $\kdf$ from a random function $R_k$ using up to $q$ queries to
  $\kdf$ and a polynomial number of operations relative to the
  security parameter $k$.
  See~\ifanonymized{{\S\,}\ref{sec:sec_impl}}\else{\cite{GuttmanEtAl2024}}\fi.
  \item[\normalfont {\scmac}:]  The {\atloc} and {\ckat} primitives
  use a Message Authentication Code $\mac$ based on a hash function
  with the collision resistance property.  The code hashing algorithm
  is also collision resistant.  We measure this by its
  \emph{existential unforgeability advantage} $\Adv{eu-mac}(A,k,q)$
  for $A$ aiming to generate a correct {\scmac} or code hash for a
  bitstring that was not queried.
  \item[\normalfont Encryption:]  The {\protfor} and {\retrfrom}
  primitives use an $\henc$ secure against chosen ciphertext
  attacks~\cite{naor1990public}.  We measure it by two values.  The
  \emph{chosen ciphertext indistinguishability advantage}
  $\Adv{indCCA2}(A,k,q)$ measures $A$'s ability to distinguish
  encryptions of two different plaintexts, and the \emph{ciphertext
    unforgeability advantage} $\Adv{eu-enc}(A,k,q)$ measures $A$'s
  ability to generate a value that will decrypt under an unknown key.
\end{description}
Code hashing for services should be second preimage resistant.

{\caif} devices thus form families, 
parameterized by a security parameter $k$.
\ifnosecimpl{The appendix}
\else
  {{\S\,}\ref{sec:sec_impl}}
\fi
argues that as $k$ increases and the advantages just described
decrease, {\caif} devices become indistinguishable from instances of
$\IF$.

\ifnosecimpl{
    \begin{corsep}\label{cor:caif:if:indistinguishable}
      If a {\caif} device has pseudorandom $\kdf$, collision-resistant
      $\mac$ and code hashing, and IND-CCA2 $\henc$, then that {\caif}
      device is computationally indistinguishable from an instance of
      $\IF$. \hfill ///
    \end{corsep}

}
\fi 

\paragraph{Hardware implementation.}
%
A hardware implementation of {\caif} is under development, with
{\fpga}s for convenience.  \textsc{asic}s will subsequently be needed,
e.g.~to protect $\IS$ properly.

The {\caif} ``special instructions'' are implemented not as
instructions, but as stores to a memory-mapped peripheral region of
the {\fpga} followed by loads from it.  FIFOs ensure that the
service's view of the process is atomic, i.e. that none of the
ciphertext for {\protfor} can be observed until all of the plaintext
has been committed.

An open-source \textsc{risc-v} soft core provides the instruction set
for the processor functionality.  Keystone~\cite{Keystone20} suggests
a provisional way to enforce the memory protection in
{\S\,}\ref{sec:caif:services}, items 1--3.  A more complete {\caif}
implementation will provide memory protection assurance directly in
hardware, using a simple layout of services in physical memory.



\section{CAIF securely implements IF}
\label{sec:sec_impl}\label{sec:properties}

We next prove that a {\caif} device is close in behavior to the ideal
functionality {\IF}, where \emph{close} is quantified by the
cryptographic properties of the primitives $\kdf$, $\mac$,
$(\henc,\hdec)$ used in the construction, as defined in
{\S\,}\ref{sec:caif:devices}.

\paragraph{Oracles.}  A computational process $A^{\mathcal{F}}$ may
make queries to a state-based process ${\mathcal{F}}$, receiving
responses from it.  We refer to the latter as an \emph{oracle}.

In any \emph{state} any \emph{query} determines a distribution
$\mathcal{D}$ on \emph{(next-state,result)} pairs.  A \emph{behavior}
or \emph{history} is a sequence of \emph{query-result} pairs, such
that there is a sequence of states where the first state is an
\emph{initial state} and, for each successive state $s$, that state
and the result are in the $\mathcal{D}$-support for the previous state
and the current query.  More formally:
\begin{definition}
  \label{defn:oracle}
  An \emph{oracle} is a tuple
  $\mathcal{O}=\seq{\Sigma, Q, I, R, \delta}$ such that
  $I\subseteq\Sigma$, $\bot\in R$, and
  $\delta\colon\Sigma\times Q\rightarrow \mathcal{D}(\Sigma\times R)$.

  $\Sigma$ is the set of states, $I$ being the initial subset.  $Q$ is
  the space of possible queries.  The function $\delta$ is the
  probabilistic transition relation.
  An alternating finite sequence
  \begin{equation}
    \label{eq:trace}
    \seq{(\sigma_0,\bot),q_0,(\sigma_1,r_1),q_1,\ldots,q_i(\sigma_{i+1},r_{i+1})}
  \end{equation}
  is a \emph{trace of} $\mathcal{O}$ iff $\sigma_0\in I$, and for all
  $j<i$,
  \begin{enumerate}
    \item $q_j\in Q$, $\sigma_{j+1}\in\Sigma$, and $r_{j+1}\in R$; and
    \item $(\sigma_{j+1},r_{j+1})\in\kind{supp}(\delta(\sigma_j,q_j))$.
  \end{enumerate}
  The probability of any trace of $\mathcal{O}$ is determined by the
  Markov chain condition, i.e.~the product of the (non-zero)
  probabilities of the successive $(\sigma_{j+1},r_{j+1})$ in the
  distributions $\delta(\sigma_j,q_j)$.

  A \emph{behavior} or \emph{history of} $\mathcal{O}$ is a finite
  sequence of pairs $\seq{\,(q_j,r_{j+1})\,}_{j<i}$ such that for some
  sequence $\seq{\sigma_j}_{j\le i}$, the sequence~(\ref{eq:trace}) is
  a trace of $\mathcal{O}$.  \hfill /// 
\end{definition}
\medskip\par\noindent We use \emph{behavior} and \emph{history} interchangeably.

$A^{\mathcal{F}(s_o,\cdot)}$ denotes running of $A$ with oracle access
to $\mathcal{F}$ with initial state $s_0$.  We write
$A^{\mathcal{F}(\cdot)}$ if the initial state is clear.
${\mathcal{F}(\cdot)}$ is function-like, returning results $r_j$ for
queries $(c_j,P_j)$.

A partial run of $A^{\mathcal{F}(\cdot)}$ induces a history
$\mathcal{H}$ of ${\mathcal{F}(\cdot)}$.  If it completes with answer
$a$, having induced the ${\mathcal{F}(\cdot)}$-history $\mathcal{H}$,
then we write $(a,\mathcal{H})\histfrom A^{\mathcal{F}(\cdot)}$.
When we do not need $\mathcal{H}$, we write
$a\leftarrow A^{\mathcal{F}(\cdot)}$ as usual.


Where $\mathcal{H} = \seq{ (q_i,r_i) }_i$ is a history,
$\query(\mathcal{H})$ refers to the sequence $\seq{q_i}_i$.  We write
$v\in_i\mathcal{S}$ when $\mathcal{S}$ is a sequence and $v$ occurs in
the $i^{\mathrm{th}}$ position of $\mathcal{S}$.
%

\subsection{Defining 
  advantages}
\label{sec:properties:defining}

A {\caif} device is an oracle with unchanging state, the intrinsic
secret $\IS$, with constant length $\length\IS$.  Its commands are the
four instructions of {\S\,}\ref{sec:caif:instructions}, syntactic
objects consisting of an instruction name together with values for the
arguments.  The result $\bot$ signals instruction failure.

The {\caif} functionalities form a family $\{C\}_k$ of oracles
parameterized by the security parameter $k$, with cryptographic
primitives yielding advantages defined in
{\S\,}\ref{sec:caif:devices}.

An instance of the {\IF} is also an oracle.  Its state consists of
{\ainstrlog} and {\ainstrescrstore}.  The transition relation $\delta$
is defined in {\S\,}\ref{sec:ideal:lemmas} using $F_{\mathit{log}}$
and the family $D_{\ell, P_s, P_r}$.  We identify the instruction
{\atloc} with {\ainstrlogval},
and so forth.

Our {\caif} and {\IF} oracles take queries $(c,p)$, where $c$ is a
syntactic command and $p$ is a service principal (i.e.~a service
hash).  As in {\S\,}\ref{sec:ideal}, we write behaviors in the form
$\seq{\,(c_j,P_j,r_j)\,}_{j<i}$ rather than
$\seq{\,((c_j,P_j),r_{j+1})\,}_{j<i}$.

$A_{q,t}$ denotes the adversary $A$ instrumented to halt immediately
if it exceeds $q$ oracle queries or $t$ computational steps.
Adversaries are possibly stateful, and may operate in phases;
$A_{q,t}$ causes all phases of $A$ combined make at most $q$ queries
and at most $t$ computational steps before halting.

\begin{definition}[{\caif} properties]
  \label{defn:caif:properties}
  Let $\{C\}_k$ be a CAIF functionality and $k$ a security parameter
  for which $C_k$ is defined.

  \begin{enumerate}    
%
    \item The \emph{attestation unforgeability advantage} of $A$ is:
    \begin{eqnarray*}
      && \hspace{-1cm} \Adv{a-u}(A,C,k,q,t) \\
      & = &
            \Pr[((p, v, \tau) ;\mathcal{H}) \histfrom A_{q,t}^{C_k(\cdot)}:\\
      & & \quad\exists i\qdot \forall j<i\qdot (\icheck(p,v,\tau),p',\true)\in_i\mathcal{H} \land \\
      & & \qquad (\iattest(v),p,\tau)\notin_j\mathcal{H}] 
    \end{eqnarray*}
    \item The \emph{protection unforgeability advantage} of $A$ is:
    \begin{eqnarray*}
      && \hspace{-1cm}\Adv{p-u}(A,C,k,q,t) \\ 
      & = &
            \Pr[((p, v, \eta) ;\mathcal{H}) \histfrom A_{q,t}^{C_k(\cdot)}:\\
      & & \quad\exists i\qdot \forall j<i\qdot (\iretrieve(p,\eta),p',v)\in_i\mathcal{H} \land \\
      & & \qquad (\iprotect(v,p'),p,\eta)\notin_j\mathcal{H}] 
    \end{eqnarray*}
    \item Let $I^{(\cdot)}(m,P_s,P_r)$ query 
    $(\iprotect(P_r,m),P_s)$ with result $\eta$.  The \emph{protection
      confidentiality advantage} of $A$ is
    $\Adv{p-c}(A,C,k,q,t) = |p_0 - p_1|$ where $p_b=$
    \begin{eqnarray*}
      & &\Pr[(m_0, m_1, P_s, P_r,\alpha) \leftarrow A_{q,t}^{C_k(\cdot)}(\bot); \\
      & &\qquad \eta\leftarrow 
          I^{C_k}(m_b,P_s,P_r); \\
      & &\qquad(x; \mathcal{H}) \histfrom
          A_{q,t}^{C_k(\cdot)}(\alpha,\eta):  \\
      & & \qquad
          x = 1 \wedge |m_0| = |m_1| \wedge \\
      &&\qquad(\iretrieve(P_s,\eta),P_r) \notin\query(\mathcal{H})]
  \end{eqnarray*}
  \end{enumerate}  
\end{definition}

\subsection{Two lemmas}
\begin{figure*}
  \centering
  \[
    \overbrace{
      \underbrace{\Pr[A^{\mathcal{O}_0(\cdot)}=1] \hspace{1.2cm} \Pr[A^{\mathcal{O}_1(\cdot)}}_
      {\small \begin{aligned}
        q^2\Adv{indCCA2}(\Gamma_{7,q}(A))
      \end{aligned}
    }
    \underbrace{=1] \hspace{1.2cm} \Pr[A^{\mathcal{O}_2(\cdot)}}_
    {\small \begin{aligned}
      \Adv{prf}(\Gamma_{8}(A))
    \end{aligned}
  }
  \underbrace{=1] \hspace{1.2cm} \Pr[A^{\mathcal{O}_3(\cdot)}=1]}_
  {\small \begin{aligned}
    q[&\Adv{a-u}(\Gamma_{9,q}(A))+\\
      &\Adv{p-u}(\Gamma_{10,q}(A))]
  \end{aligned}
}
}^{\begin{aligned}
  \Adv{ind}(A,\caif,k,q)
\end{aligned}}
\]
\caption{Implementation theorem proof strategy:
  Lemmas~\ref{lem:a-u} and~\ref{lem:p-u} bound the last term}
  \label{fig:proof:strategy}
\end{figure*}

Let $k$ be a security parameter for which {\caif} is defined.

\begin{lemma}[Attestation unforgeability]
\label{lem:a-u}
%
There are reductions $\Gamma_{1,q}$ and $\Gamma_2$ with additive
computational overhead, such that for any $A$,
$\Adv{a-u}(A,{\caif},k,q,t) \leq$
  \begin{eqnarray*}
    &&
       q \Adv{eu-mac}(\Gamma_{1,q}(A),k,q) + \\
    & &\Adv{prf}(\Gamma_2(A),k,q+1)  \qquad\qquad ///
  \end{eqnarray*}
\end{lemma}
Proof:  See\ifanonymized{~(anonymized material)
  }\else{~\cite{GuttmanEtAl2024} }\fi for proofs.

\begin{lemma}[Protection unforgeability]
  \label{lem:p-u}
  There are reductions $\Gamma_{3,q}$ and $\Gamma_4$ with additive
  computational overhead $t_3(k,q)$ and $t_4(k,q)$ respectively, such
  that for any $A$, $\Adv{p-u}(A,{\caif},k,q,t)\leq$
  \begin{eqnarray*}    
    && q \Adv{eu-enc}(\Gamma_{3,q}(A),\mathcal{E},k,q,t+t_3(k,q)) + \\
    & &\Adv{prf}(\Gamma_4(A),\kdf,k,q+1,t+t_4(k,q))  \qquad\qquad ///
  \end{eqnarray*}
\end{lemma}

\subsection{Proving secure implementation}


Given a {\caif} device with its crypto primitives and a particular
intrinsic secret $\IS$, let $\IF$ be the instance
{\IF}$[F_{\mathit{log}}, \{D_{\ell, P_s, P_r}\}]$ where:
\begin{description}
  \item[$F_{\mathit{log}}(v,P)$] $=\mac(sk,v)$ 
  where $sk=\kdf(\mbox{\texttt{"at"}}, \; \IS, \; P)$;
  \item[$D_{\ell, P_s, P_r}$] is the distribution generated by
  $\henc(0^\ell,sk)$  \\
  where $sk=\kdf(\mbox{\texttt{"pf"}}, \; \IS, \; P_s, \;P_r)$.
\end{description}

\begin{theorem}\label{thm:securely:implements}
  Let $\{C_k\}$ be a CAIF functionality and let $k$ be a security
  parameter for which which $C$ is defined.  Let $\Adv{imp}(A,C,k,q,t)$
  be defined to be
  $$|\Pr[A^{C_k(\cdot)}=1] - \Pr[A^{\IF_k(\cdot)}=1]|.$$
  There are reductions $\Gamma_{7,q}$, $\Gamma_{8}$, $\Gamma_{9,q}$,
  and $\Gamma_{10,q}$ with computational overhead times $t_7(k,q)$,
  $t_8(k,q), t_9(k,q)$, and $t_{10}(k,q)$ respectively, together with
  the reductions $\Gamma_{1,q}$, $\Gamma_2$, $\Gamma_{3,q}$, and
  $\Gamma_4$ of Lemmas~\ref{lem:a-u}--\ref{lem:p-u}, s.t. for all $q <
  W$,
  \begin{align*}
    \Adv{imp}&(A,{\caif},k,q,t) \leq\\
             &q^2\Adv{indCCA2}(\Gamma_{7,q}(A),k,q) + \\ 
             &\Adv{prf}(\Gamma_{10}(A),k,q)\\
             &q^2(\Adv{eu-mac}(\Gamma_{1,q}(\Gamma_{9,q}(A)),k,q)) +\\
             &q(\Adv{prf}(\Gamma_2(\Gamma_{9,q}(A)),k,q+1)) \\ 
             &q^2(\Adv{eu-enc}(\Gamma_{3,q}(\Gamma_{10,q}(A)),k,q)) +\\
             &q(\Adv{prf}(\Gamma_4(\Gamma_{10,q}(A)),k,q+1)) 
  \end{align*}
\end{theorem}

\begin{proofsketch}
  The proof operates as a hybrid argument regarding four
  probabilities, namely the probabilities that
  $A^{\mathcal{O}_i(\cdot)}$ outputs 1 for various oracles
  $\mathcal{O}_i$ for $0 \leq i \leq 3$.  The oracles are defined as
  follows:

  \begin{description}
    \item{$\mathcal{O}_0$} $=\IF$.
    \item{$\mathcal{O}_1$} implements $\IF$, except that for $\aprotect$
    queries, we select the candidate values $c$ by encrypting the
    input value $v$ rather than $0^{|v|}$.
    \item{$\mathcal{O}_2$} implements $\mathcal{O}_1$ except that it
    uses $\kdf$ with an intrinsic secret $\IS$ instead of using the
    random function $R_k$.
    \item{$\mathcal{O}_3$} $= C_k$.
  \end{description}
We construct a reduction for each gap, summarized in
Fig.~\ref{fig:proof:strategy}.  Lemmas~\ref{lem:a-u}--\ref{lem:p-u}
bound the rightmost term in Fig.~\ref{fig:proof:strategy}.  \hfill ///

\end{proofsketch}

Hence, if {\caif} uses strong cryptography:
%
\smallskip
\begin{cor}\label{cor:caif:if:indistinguishable}
  If a {\caif} device has pseudorandom $\kdf$, existentially
  unforgeable $\mac$, collision-resistant code hashing, and IND-CCA2
  $\henc$, then that {\caif} device is computationally
  indistinguishable from an instance of $\IF$. \hfill ///
\end{cor}
%

\section{Assured Remote Execution Strategy}
\label{sec:strategy}
In {\S\S\,}\ref{sec:strategy}--\ref{sec:delegation} we verify
protocols for assured remote execution on {\caif} against a
code-selecting adversary.  Each device has a uniquely identifying
name, its \emph{immutable identifier} $\id$; $\id$s, like other names,
may be publicly known.

When we speak of \emph{protecting} a value \emph{for} a service or
\emph{retrieving} a value \emph{from} a service, we mean that the
value is the argument to {\protfor} or the result of {\retrfrom},
resp.

\subsection{Achieving Assured Remote Execution}
\label{sec:strategy:achieve:arex}

Our strategy uses a succession of steps, 
and does not depend on any digital signature algorithm to remain
secure throughout a device's lifetime.
%
%
%
We start from an assumption.  The device must be started running a
known program once, at the beginning of its lifetime.  The
manufacturer---already trusted to produce correct
hardware---initializes it with a compliant service to run first.  This
\emph{anchor} service must run \emph{only} in a secure environment.
Devices undergo a state change---such as fusing a wire, flipping a
switch, or advancing a monotonic counter---to prevent re-execution
after successful anchoring.

The owner or authority controlling this device, its \emph{device
  authority} {\da}, anchors the device, sharing the secret $k_s$ with
it in step~\ref{item:strategy:anchor}.  The {\da} must store $k_s$
securely for later use.

Starting with our assumption: 
\begin{renumerate}\setcounter{enumi}{-1}
  \item We assume \emph{local} execution of a single service, the
  \emph{anchor service} in a context suitable for secure
  initialization.\label{item:strategy:assume}
  
  \item \label{item:strategy:anchor}
  The {\da} and the anchor service establish a shared secret $k_s$.
  The anchor service protects $k_s$ for the exclusive use of a
  recipient service $\svc_1$ via {\protfor} in shared storage.  It
  then zeroes its unshared memory and exits
  ({\S\,}{\ref{sec:anchoring}}).

  A state change prevents rerunning the anchor service.

  \item \label{item:strategy:distributor} $\svc_1$ is a
  \emph{symmetric key distributor service}.  It will receive
  authenticated requests from {\da}; each request specifies a service
  hash $\shash$, and $\svc_1$ derives a secret
  $\kind{kdf}(k_s,\shash)$ that it protects for the exclusive use of
  the service with hash $\shash$ in shared storage.  It then zeroes
  its unshared memory and exits ({\S\,}{\ref{sec:anchoring:distr}}).

\end{renumerate}
For instance, the key $\kar$ of {\S\,}\ref{sec:using:techniques} is
derived by the symmetric key distributor service as
$\kar=\kind{kdf}(k_s,\oh)$.

The anchor service and symmetric key distributor are not replaceable.
The anchor service is run only at the time of acquisition.  The
distributor is used only for one simple function, meaning that it can
be programmed correctly; moreover, it uses only symmetric
cryptography, which is much stabler and longer lasting then asymmetric
algorithms may be.

To escape from using shared secrets to infer assured remote execution,
we establish a \emph{signing key delegation service}.  Possibly long
after the time of initialization, possibly repeatedly,
{\S\,}{\ref{sec:delegation}} installs new programs and authorizes them
using a key from the symmetric key distributor:
%
%
\begin{renumerate}
  \item \label{item:strategy:set:up} A \emph{set-up service} with hash
  $\setupdelhash$ generates a device signature key pair $(\dk,\dvk)$.
  Using the shared secret $\kind{kdf}(k_s,\setupdelhash)$ to
  authenticate, it proves to a certifying authority $\ca$ operated by
  {\da} that it holds the signing key $\dk$.  A certificate associates
  the verification key $\dvk$ with a service hash $\delhash$, the
  device's $\id$, and some supplemental values.
  %
%
%

  The set-up service protects $\dk$ for service $\delhash$
  exclusively.  It then zeros its unshared memory, and exits.
  \item \label{item:strategy:delegate} The \emph{delegation service}
  has hash $\delhash$.  When invoked with a target service hash
  $\shash$, it generates a new signing key pair $(\sk,\vk)$.  It emits
  a certificate-like binding
  $\tagged{\ldots\id,\shash,\vk,\ldots}{\dk}$ signed with $\dk$.

  It protects $\sk$ for the service $\shash$ exclusively, in shared
  memory.  It then zeros its unshared memory, and exits.
\end{renumerate}
If $\shash$ does not expose $\dk$, then messages signed with $\sk$
must come from $\shash$, as required for assured remote execution.
  
\paragraph{Assured remote execution.}
Each of these steps builds additional assured remote execution power.
Step~\ref{item:strategy:assume} simply assumes local execution of the
anchor service in secure initial environment.
Step~\ref{item:strategy:anchor} assures remote execution only for
\emph{one} service $\svc_1$, with evidence useful only to {\da}, which
holds $k_s$.
Step~\ref{item:strategy:distributor} gives \emph{any} service $\shash$
a secret, but evidence of $\shash$'s remote execution is useful only
those sharing the secret.

Step~\ref{item:strategy:delegate} completes the progression.
\emph{Any} $\shash$ can receive a signing key documenting its
execution on $d$, and any principal willing to trust the $\ca$'s
certificate can use the evidence.  No shared secrets are needed as
$\shash$ executes.



\paragraph{Alternative for short-lived devices.}  If $d$ will never
outlive a signing algorithm and keysize, a simpler protocol is enough.
The set-up service of step~\ref{item:strategy:set:up} becomes the
anchor service.  It obtains a certificate on $\dvk$.  The signing key
delegation service of step~\ref{item:strategy:delegate} provides
signing keys to services $\shash$.  No shared secrets are needed;
however, if algorithms or root keys are superseded, assured remote
execution cannot be reestablished.  Other benefits of local flow control
remain.

\subsection{Compliant roles and adversary roles}
\label{sec:adversary:compliant:adversary}
\label{sec:adversary:adversary}
\label{sec:adversary:formalizing}

Over time, many active processes will execute and use the {\caif}
instructions on a device.  Many of the processes may behave
unpredictably, either because the adversary chose them, or because
they were simply poorly written.  They may use the {\caif}
instructions with any values they can obtain; {\caif} just ensures
they do so under their own service hash.  We call them
\emph{adversarial services} or just \emph{wildcats}.

However, some programs, having been carefully constructed for specific
behaviors such as the ones we have just described, may have no other
relevant behaviors.  They do not use these keys for any other
messages, nor use the {\caif} instructions in any other relevant way.
We call them \emph{compliant services}.

A compliant service complies with one or more \emph{role
  specifications}, and never performs relevant actions except as the
role dictates, even under adversarial inputs that may seek to exercise
Return-Oriented Programming gadgets (e.g.).  The predicate {\regular}
is true of service hashes of programs, intuitively, that we have
vetted and judge compliant.  If a service hash $\CH$ is {\regular},
the consequence is:
\begin{itemize}
  \item every use of the {\caif} instructions with active service hash
  $\CH$ belongs to an instance of a specified compliant role,
\end{itemize}
i.e.~the roles with behaviors defined in
{\S\S\,}\ref{sec:anchoring}--\ref{sec:delegation}.  The {\regular}
property always arises as an \emph{assumption} in analysis; we will
not present methods here for proving services are compliant,
presumably a task for program verification.  However, these roles have
well defined correctness conditions and lack subtle control flow and
sophisticated data structures.

Adversary activities or \emph {wildcats} use the {\caif} instructions
freely, but use service hashes not assumed {\regular}.

\paragraph{Wildcat roles.}  We specify the adversary's local
powers---beyond the usual network powers to interrupt, redirect, and
synthesize messages, to extract and retain their contents, and to
execute cryptographic operations using any keys they may possess---by
three \emph{wildcat} roles with names beginning \texttt{wc-},
so-called because they may use the {\IF} instructions in whatever
unexpected patterns would benefit the adversary:
\begin{description}
  \item[wc-protect] causes an {\iprotect} instruction with current
  service $\CH_s$, recipient $\CH_r$, and value $v$.
  \item[wc-retrieve] causes an {\iretrieve} instruction and transmits
  $v$ for adversary use.
  %
  \item[wc-attest] causes an {\iattest} instruction, logging $v$ with
  the current service $\CH_s$ in the {\ainstrlog}.
\end{description}
A wildcat role for {\ckat} is unnecessary in this context, because it
is a conditional that produces no new data.  The cases are represented
for protocol analysis via pattern matching.

\paragraph{Rules on the wildcat roles.}  The wildcat roles obey rules
saying that if $\CH$ is the active hash in a wildcat role, then $\CH$
is not {\regular}.  So {\caif} instructions with {\regular} service
hashes occur \emph{only} in rule-bound, non-wildcat roles.

This is a strong adversary model:  the adversary may install any
programs and execute them as services, using {\protfor} etc.~as
desired.  However, if a program has a service hash we assume
{\regular}, then it will have the same behavior we specified in a
compliant, non-wildcat role.  The adversary can also run our services,
unless special provisions prevent some from running, as e.g.~repeated
anchoring (step~\ref{item:strategy:anchor}).

\section{Anchoring a CAIF Device}
\label{sec:anchoring}

When a new {\caif} device reaches the buyer's warehouse---or
alternatively, just before shipping---it is \emph{anchored}.

Anchoring runs a known program on the fresh {\caif} device in a secure
environment.  This may require shielded cables to the device, or
wireless communications shielded in a Faraday cage.  Thus, we will
call this the \emph{ceremony in the metal room}.

The metal room provides \emph{authenticated} and \emph{confidential}
channels between {\da} and the device, thereby
delivering a shared secret $k_s$.  Subsequently the device and {\da}
use $k_s$ and derived keys for symmetric encryptions and {\scmac}s.

%
\label{sec:anchoring:ss}

\paragraph{The anchor service.}  The steps of anchoring are:
\begin{enumerate}
  \item {\da} turns on $d$ and starts $d$ running the \emph{symmetric
    anchor service}, with service hash $\anch$.
  \item {\da} observes the device identifier $\id$, and prepares a
  secret seed $r$ and a nonce $n$.  {\da} transmits
  \[ \id, \anch, \dhash, n, r ,  
  \]
  $\dhash$ being the hash of the destination service meant to obtain
  the secret $k_s$.
  %
  %
  \item The service $\anch$ checks its device has id $\id$, and its
  service hash is $\anch$.  It then computes
  \[ k_s=\kdf(r,\id) , \]
  and replies with  $n$ to confirm completion.  
  \item It uses {\protfor} to protect $k_s$ together with the
  identities $\seq{\dhash,\anch}$ for service $\dhash$ exclusively
  into shared storage; it zeroes its unshared memory and exits.
\end{enumerate}
Hence, $k_s$ is available on $\id$ only to the service with hash
$\dhash$.  To ease managing long-term secrets, $r$ may itself be
derived by differentiating a group seed $r_0$ for $\id$, setting
$r=\kdf(r_0,\id)$, so $k_s=\kdf(\kdf(r_0,\id),\id)$.


The anchor service must run \emph{only} with a secure metal room
channel, requiring an irreversible state change after which $\anch$
will no longer run, e.g.~a switch being flipped.

After the ceremony in the metal room, $k_s$ remains permanently
available to $\dhash$ to secure later remote interactions.

\subsection{Analyzing symmetric anchoring}
\label{sec:adversary:anchor}

We now analyze the ceremony in the metal room.
%
%

A role on the device represents the \emph{anchor} service, together
with a role \emph{dev-init-imid} to initialize a device's immutable ID
$\id$ to a fresh value.
  \begin{figure}
    \centering
    \[
      \xymatrix@R=3mm@C=4mm{ \mbox{\footnotesize dev-get-anc} &
        \mbox{\footnotesize da-send-anc} & \mbox{\footnotesize
          dev-init-imid} & \mbox{\footnotesize da-init-group-seed} \\
        & \bluenode\ar@{=>}[d] & \blacknode\ar@{=>}[dd]\ar[l]_{\id} &\graynode\ar@/^/[lld]^<<<<<{r_0} \\
        &\graynode\ar@{=>}[d] &  & \\
        \bluenode\ar@{=>}[d] & \blacknode\ar[l]_{\ldots r} & \graynode\ar@/^/[lld]_<<<<{\id} & \\
        \graynode\ar@{=>}[d] &&& \\
        \graynode\ar[rr]^{\mathsf{prot(\dhash,(k_s,\id,\seq{\anch,\dhash}))}}
        &&& }
    \]
    \caption[Symmetric anchoring]{Metal room activity, symmetric
      anchoring}
    \label{fig:anchor:analysis}
  \end{figure}
  %
%
A pair of roles represent the {\da}'s interaction with the device.
One does set-up to manage the shared secrets, creating a \emph{group
  seed} $r_0$, from which secrets $r$ are derived.
A second \emph{sends the anchor} secret $r$ during the ceremony in the
metal room.


Our analyses all assume that the anchor service hash $\anch$ is
{\regular}, i.e.~that the adversary cannot perform wildcat actions
under hash $\anch$.  We ascertain that $k_s$ and its derivatives
remain unavailable to the adversary.

What happens at the metal room layer if the device \emph{anchor} role
%
%
runs to completion?  We assume the metal room ensures both
authenticity and confidentiality.  We find that the anchor secret
$k_s$ is obtained from the {\da} sending the secret $r$.  The two
parties obtain the device's $\id$ from the same device events.  And
the {\da} has properly derived the anchor secret from its group seed
$r_0$ (Fig.~\ref{fig:anchor:analysis}).

In Fig.~\ref{fig:anchor:analysis}, vertical columns represent
successive actions---reading downwards along the double arrows---of an
instance of the role named at the top; single arrows represent
propagation of messages or of values in device-local state ({\protfor}
records, secure seed storage at the {\da}).  Message transmission and
reception are shown as black nodes $\blacknode$ and blue nodes
$\bluenode$, resp.  Local state reads and writes are gray nodes
$\graynode$.  More comments on the diagrams are at the end of
{\S\,}\ref{sec:delegation:adapting}.

In subsequent steps, as we add more roles to model subsequent
activities we reverify these properties, since protocol interactions
could undermine them.

\subsection{Trust chains}
\label{sec:anchoring:trust:chains}

We keep track of ``chains of provenance'' for trust, meaning the
sequence of services---starting from an anchor service---that obtained
previous keys and generated new keys to protect for their successors.
We store trust chains with the keys that they validate in {\protfor}
records.  We also deliver trust chains in messages between parties,
who can check them to trace provenance back to the anchoring program.

Trust chains are lists of service hashes.  If $\trustchain$ is a trust
chain, we write the effect of pushing a new hash $h$ to it as
$h\cons\trustchain$, using a \emph{cons} ``${\cons}$'' operation.
%

Each of the services we specify, when retrieving a trust chain, tests
that its code hash is the first and the source service's hash is
second.  Thus, it has confirmed by the successful {\retrfrom} that the
prior entry also represented its identity correctly.  When extending a
trust chain, it pushes its intended recipient's hash to the front of
the list.

If an observer, seeing the trust chain, knows that the front $n$
entries are \emph{compliant} roles we have specified, it follows that
the $n+1$st entry is the actual source of the record retrieved by the
$n$th entry.  If that hash is also known to be compliant, this process
can repeat.

We have successfully designed and verified protocols relying on trust
chains of length up to five (see {\S\,}\ref{sec:delegation}).

\subsection{The symmetric key distributor service}
\label{sec:anchoring:distr}


The anchor service may be used with any $\dhash$.  A useful $\dhash$
is a \emph{symmetric key distributor service}, whose hash we will
write $\dstrh$.  The symmetric key distributor program retrieves $k_s$
packaged with the trust chain
$\trustchain=\dstrh\cons\anch\cons\seq{}$, checking its hash $\dstrh$
and its source $\anch$.  It then uses $k_s$ to decrypt a message from
the {\da}.  This message should contain:
\begin{itemize}
  \item a target service hash $\acth$ for which a new symmetric key
  $k$ should be derived;
  \item a trust chain $\trustchain'
  $;
  \item a payload $\payload$ to pass to the service $\acth$ when it
  runs.
\end{itemize}
The distributor service checks that $\trustchain'=\trustchain$.  If
so, $\dstrh$ derives the key $k=\kdf(k_s,\acth)$.  It protects an
extended trust chain, the payload, and this key $k$ for $\acth$:
\[ {\protfor}\; \acth\; (\id, \; (\acth\cons\trustchain), \; \payload,
  \; k)
\]
%
into shared storage, after which it zeroes its unshared memory and
exits.  On device $\id$, $k$ can only by obtained by the service
$\acth$.  Since the {\da} can compute $k=\kdf(k_s,\acth)$, $k$ enables
the {\da} to create an authenticated, confidential channel to $\id$
where only service $\acth$ can be the peer.


\subsection{Analyzing symmetric key distribution}

This layer of analysis has a role representing the \emph{symmetric key
  distributor} service on the device, together with a role
\emph{use-it} that retrieves the distributor's key and uses it
generate a confirmation message.  A third role on the {\da} delivers
the \emph{distribution request} to the device.  It terminates
successfully after receiving a confirmation message from $\acth$.

The analysis for {\regular} code hashes for $\anch$, $\dstrh$, and the
target service $\acth$ appears in Fig.~\ref{fig:dtr:req:analysis}.
The distribution service uses $k_s$ generated from $r_0$ given in
Fig.~\ref{fig:anchor:analysis}; it handles the request and derives
$k$, protecting it for $\acth$.  The latter uses $k$ to prepare a
confirmation for the requester.  The value $\mathit{seed}$ is stored
in state in the {\da}, while the value $k_s$ is stored in the device
state inside a \emph{protect-for} record, generated in the lower left
node of Fig.~\ref{fig:anchor:analysis}.
\begin{figure}
  \centering
  \[
    \xymatrix@C=12mm@R=3mm{*+[F]{\strut \txt{\footnotesize Metal room
          activity
          \\
          \footnotesize from Fig.~\ref{fig:anchor:analysis}}}
      \ar[rd]|<<<<<<<<{r_0,\id}
      \ar[rrrd]|>>>>>>>>>>>>>>>{\;k_s,\id,\ldots\;} & 
      \mbox{\footnotesize dstr-req} & \mbox{\footnotesize use-it} &
      \mbox{\footnotesize dstr} &
      \\
      & \graynode\ar@{=>}[d] & & \graynode\ar@{=>}[d] & \\
      & \blacknode\ar@{=>}[dd]\ar[rr] & & \bluenode\ar@{=>}[d] & \\
      & & \graynode\ar@{=>}[d] & \graynode\ar[l]_{\ldots k\ldots} & \\
      & \bluenode &\blacknode\ar[l] & & }
  \]
  \caption{Distributor request}
  \label{fig:dtr:req:analysis}
\end{figure}

If $\acth$ is not assumed {\regular}, a separate analysis shows that,
as an alternative to Fig.~\ref{fig:dtr:req:analysis},
\texttt{wc-retrieve} can expose $k$.  This is correct in our strong
adversary model.

Because the arrows in Fig.~\ref{fig:dtr:req:analysis} point out of the
box at the upper right, the behaviors in the remainder of the diagram
must come after the anchoring ceremony.  But because no arrows point
\emph{into} the box at the upper right, no part of the remainder has
to come before the anchoring ceremony.  Assuming that the values
stored in state records persist, the distributor may be used to set up
keys for software long after the ceremony in the metal room, e.g.~when
the device is in orbit on a satellite.  This functionality assures
long-term remote execution without local contact.

\section{Delegating Signing Keys}
\label{sec:delegation}

A \emph{delegation service} holding a signing key $\dk$ behaves as
follows to generate a signing key for a service $\shash$:
\begin{enumerate}
  \item It generates a fresh target signature key pair $(\sk,\vk)$;
  \item It emits a certificate-like message signed with $\dk$ that
  binds the new verification key $\vk$ to $\shash$ and the device
  $\id$\ifnosecimpl{ }\else{ and additional information}\fi.  We
  call this a \emph{delegation certificate}.
  \item It protects the signing key $\sk$ and additional information
  for the exclusive use of service $\shash$, placing it in shared
  storage.  It then zeroes its unshared memory and exits.
\end{enumerate}
If a recipient knows the delegation service generated the delegation
certificate and observes a message signed with $\sk$, it can infer
$\shash$ has been active on $\id$.  To ensure the delegation service
generated the certificate, we (in turn) use a \emph{delegation set-up
  service} to get a $\ca$ certificate for $\dk$.  Call the delegation
service's hash $\delhash$ and the set-up service's hash
$\setupdelhash$.

The delegation set-up service acts as follows:
\begin{enumerate}
  \item It obtains a \emph{certify request} ultimately from the $\ca$
  containing a fresh $\serial$ number for the resulting certificate;
  \item It generates a signing key pair $(\dk,\dvk)$ for $\delhash$;
  \item It transmits a proof-of-possession message using the signing
  key $\dk$ to the $\ca$ on an authenticated channel;
  \item It receives a  certificate; and 
  \item It protects $\dk$ together with additional information for the
  sole use of $\delhash$ into shared storage; it zeroes its unshared
  memory and exits.
\end{enumerate}
The additional information mentioned includes trust chains asserting
full data provenance ({\S\,}\ref{sec:anchoring:trust:chains}).

\ifnosecimpl{\vspace{-1ex}}\else{}\fi
\subsection{Assumptions and security goal for delegation}
\label{sec:delegation:assumptions}

Our delegation scheme relies on the assumptions:
\begin{enumerate}
  \item [(i)] the $\ca$ is uncompromised;
  \item[(ii)] the $\ca$ receives the proof-of-possession from
  $\setupdelhash$ on $\id$ on an authenticated channel when certifying
  $\dvk$.\label{item:the:assumptions}
\end{enumerate}
The anchoring in {\S\,}\ref{sec:anchoring} will provide the
authenticated channel required in Assumption~(ii).
%
Moreover, for a particular target service with service hash $\shash$ one
may assume:
\begin{enumerate}
  \item[(iii)] the service $\shash$ uses $\sk$ for signing messages,
  but not in any other way; hence $\shash$ does not disclose $\sk$.
\end{enumerate}
%

\label{sec:delegation:goal}

Suppose one observes:  a digital certificate $m_1$ from $\ca$ binding
$\dvk$ to $\delhash$ on $\id$; a delegation certificate $m_2$ binding
$\vk$ with $\shash$ on $\id$; and a message $m_3=\tagged{m_0}{\sk}$
verifying under $\vk$.

Then our \emph{main security goal} is:  When assumptions (i) and (ii)
hold, and (iii) holds for this $\shash$, then, on device $\id$:
\begin{enumerate}
  \item The delegation set-up service has generated $\dk$, obtained
  the certificate $m_1$ on $\dvk$ and $\delhash$, and protected $\dk$
  solely for the delegation service $\delhash$;
  \item The delegation service $\delhash$ generated $(\sk,\vk)$,
  emitted the certificate $m_2$ on $\shash$ and $\vk$, and protected
  $\sk$ for the sole use of the service $\shash$;
  \item This service $\shash$ used $\sk$ to sign $m_0$, yielding
  $m_3$.
\end{enumerate}
The observer thus infers, subject to (i)--(iii), that the delegation
process proceeded correctly, and that $\shash$ is responsible for
$m_0$.  This is the assured remote execution claim for $m_0$.

\subsection{Message forms}
\label{sec:delegation:msgs}

We use tags to distinguish tuples of components that might be
confused.  Here, we only show the components, not the tags.  The
\emph{certification request} contains the components:
\[ \id, \delhash, \setupdelhash, \trustchain, \serial, \ca , \]
where $\trustchain$ is a trust chain acceptable to the $\ca$.  The
proof-of-possession is signed with the signing part of the key pair
and contains the verification key:
\[ \tagged{\serial, \id, \delhash, \setupdelhash, \trustchain,
    \dvk}{\dk} ; 
\]
the $\ca$ must ascertain that it arrives on an authenticated channel
from service $\setupdelhash$ running on $\id$.  The anchoring enables
this.  The resulting certificate is of the form:
\[ \tagged{\id, \delhash, \setupdelhash, \trustchain, \serial,
    \dvk}{\ca} .
\]
The set-up service protects $(\dk,\dvk)$ plus an expanded trust chain
$\trustchain'=\delhash\cons \setupdelhash\cons\trustchain$ via
$\protfor$.  The delegation service then uses $\retrfrom$ to recover
the key pair $(\dk,\dvk)$ and $\trustchain'$.  When generating a key
pair $(\sk,\vk)$ for a target service $\shash$, it protects
$\shash\cons\trustchain'$ and $(\sk,\vk)$ for $\shash$ with
$\protfor$, and emits a delegation certificate
\[ \tagged{\id, \shash, \trustchain', n, \vk}{\dk} .
\]

\ifnosecimpl{\vspace{-1ex}}\else{}\fi
\subsection{Delegation analysis}
\label{sec:delegation:analysis}

%
%
%
\begin{figure} 
  \centering
  \[ \xymatrix@R=2mm@C=5mm{ 
      &&&\ca & \mbox{\small setup} & \mbox{\small del} &
      \mbox{\small\emph{sh}}
      \\
      &&& \bullet\ar@{=>}[dd]\ar@{-->}[r] & \circ\ar@{=>}[d] && \\
      &&& & \graynode\ar@{=>}[d]\ar[r] & \graynode\ar@{=>}[d] & \\
      m_1 & m_2 & m_3 & \bluenode\ar@{=>}[d] & \blacknode\ar[l]^{\textcolor{darkgreen}\star}
      &\graynode\ar@{=>}[d]\ar[dr]& \\
      \bluenode &&& \blacknode\ar[lll] & & \blacknode\ar@/^/[dllll] &
      \graynode\ar@{=>}[d]\\
      &\bluenode & &&&& \blacknode\ar@/^/[dllll] \\
      && \bluenode &&&&}
  \]
%
  \caption{Message $m_3$ signed with delegated key;
    ${\textcolor{darkgreen}\star}$ authenticated channel}
  \label{fig:delegated:key:shape}
\end{figure}
We analyze the delegation mechanism over a generic authenticated
channel by querying what must have occurred if the following three
messages are observed:
\begin{description}
  \item[\normalfont$m_1$:] a certificate
  $\tagged{\id, \delhash, \setupdelhash, \trustchain, \serial,
    \dvk}{\ca}$ for the delegation verification key $\dvk$;
  \item[\normalfont$m_2$:] a delegation certificate
  $\tagged{\id, \shash, \trustchain', n, \vk}{\dk}$ for the service
  $\mathop{sh}$'s key $\vk$;
  and
  \item[\normalfont$m_3$:] a message $\tagged{m_0}{\sk}$ in which
  $m_0$ is signed by some $\sk$ forming a key pair $(\sk,\vk)$ with
  the certified $\vk$.
\end{description}
These messages are in the three left columns of
Fig.~\ref{fig:delegated:key:shape}.

Fig.~\ref{fig:delegated:key:shape} shows the single result of
{\cpsa}'s analysis under the assumptions (i)--(iii) from
{\S\,}\ref{sec:delegation:assumptions}.  {\cpsa} determines this is
the only way that the certificates and signed message
$\tagged{m_0}{\sk}$ can be observed, subject to (i)--(iii).  The white
node $\circ$ and dashed line will be expanded subsequently (see
Fig.~\ref{fig:delegated:key:shape:anchored}).

In this scenario, $\tagged{m_0}{\sk}$ was in fact generated by the
expected (rightmost) role instance $\mathop{sh}$, which obtained its
key from the delegation service to its left; that in turn generated
the certificate on $\vk$ and obtained its signing key $\dk$ from the
delegation set-up service preceding it, which in fact interacted with
the $\ca$ to generate the certificate.

\ifnosecimpl{\vspace{-1ex}}\else{}\fi
\paragraph{Assured remote execution.}
Fig.~\ref{fig:delegated:key:shape} shows the assured remote
execution guarantee, subject to assumptions (i)--(iii).  Messages
signed by $\sk$ come from the program $\shash$ on device $\id$.

As usual, we can build authenticated and confidential channels to
$\shash$ on top of this, e.g.~using $m_0$ to send a public key for a
Key Encapsulation Mechanism~\cite{shoup2001proposal}.

\subsection{Adapting delegation to the anchor}
\label{sec:delegation:adapting}

Anchoring provides a concrete realization of the authenticated channel
for the proof-of-possession.  The symmetric key distributor
({\S\,}\ref{sec:anchoring:distr}) generates
$k_{\mathop{sud}}=\kdf(k_s,\setupdelhash)$; $\setupdelhash$ uses it to
encrypt the proof-of-possession to the $\ca$.

The symmetric key distributor receives a payload $\payload$ from the
{\da} when deriving a key, which it protects for the recipient with
the key.  We use the {$\ca$}'s certify request
$\id, \delhash, \setupdelhash, \trustchain, \serial, \ca$ as this
$\payload$.  The delegation set-up service retrieves this using
{\retrfrom}, together with a trust chain $\trustchain_1$ and the
derived key $k_{\mathop{sud}}$.  The trust chain $\trustchain$ is the
{\da}'s acceptable chain of custody for $k_{\mathop{sud}}$, while
$\trustchain_1$ contains its actual history, as analysis confirms.
The set-up service does not proceed unless
$\trustchain=\trustchain_1$.

Since $\anch$, $\dstrh$, $\setupdelhash$, and $\delhash$ are all in
$\trustchain$, the $\ca$ emits a certificate only when the trust chain
is acceptable.  We consider the case that they are all {\regular}.

If the target service hash $\shash$ is {\regular}, analysis yields a
single possibility shown in
Fig.~\ref{fig:delegated:key:shape:anchored} that enriches the
Fig.~\ref{fig:delegated:key:shape}.  The proof-of-possession
$\mathop{pop}$ is encrypted under $k_{\mathop{sud}}$ on the
${\textcolor{darkgreen}{\star\star}}$ arrow from $\setupdelhash$ to
$\ca$.  If $\shash$ may not be {\regular}, a wildcat retrieve role may
alternatively extract and disclose $\sk$, instead of $\shash$'s run,
as expected.
%

\ifnosecimpl{\vspace{-1ex}}\else{}\fi
\paragraph{About the diagrams.}  We have redrawn and simplified
{\cpsa}'s diagrams.  We reordered role instances for clarity.  We
grouped copies of Fig.~\ref{fig:anchor:analysis} in a single box.  We
combined adjacent state nodes that are distinct for {\cpsa}.
%
%
%
{\cpsa} also emits information about messages transmitted and
received, etc., mentioned here only in our accompanying text.
\begin{figure}
  \centering
  \[ \xymatrix@R=2mm@C=2mm{ 
      *+[F]{\strut \txt{\tiny Metal room
          \\
          \tiny (Fig.~\ref{fig:anchor:analysis})}}
      \ar@/_/[rrrd]|{r_0,\id}\ar[rrrrrrrd]|{\;k_s,\id,\ldots\;} &&&\ca &
      \mbox{\small setup} & \mbox{\small del} & \mbox{\small\emph{sh}}
      & \mbox{\small dstr}
      \\
      &&&\graynode\ar@{=>}[d]&&& & \graynode\ar@{=>}[d]  \\
      &&&\bullet\ar@{=>}[ddd]\ar[rrrr]   &&&& \bluenode\ar@{=>}[d]  \\
      &&&  & \graynode\ar@{=>}[d] &&& \graynode\ar[lll] \\
      &&& & \graynode\ar@{=>}[d]\ar[r] & \graynode\ar@{=>}[d] & \\
      m_1 & m_2 & m_3 & \bluenode\ar@{=>}[d] &
      \blacknode\ar[l]^{\textcolor{darkgreen}{\star\star}}
      &\graynode\ar@{=>}[d]\ar[dr]& \\
      \bluenode &&& \blacknode\ar[lll] & & \blacknode\ar@/^/[dllll] &
      \graynode\ar@{=>}[d]\\
      &\bluenode & &&&& \blacknode\ar@/^/[dllll] \\
      && \bluenode &&&}
  \]
  \caption[Anchored delegated key usage]{Message $m_3$ signed, with
    anchoring and ${\textcolor{darkgreen}{\star\star}}$ encrypted}
  \label{fig:delegated:key:shape:anchored}
\end{figure}

\section{Related Work}
\label{sec:related}

%


\subsection{Trusted Execution Environments}
\label{sec:related:tees}
\nocite{schneider2022sok}
\nocite{IntelTDX,cheng2023intel}
\nocite{cryptoeprint:2016:086,GuttmanRamsdell19}
\nocite{Keystone20}
\nocite{MunozRRL23}

{\caif} is a kind of {\tee}, like Intel's {\textsc{sgx}} and
{\textsc{tdx}}~\cite{cryptoeprint:2016:086,IntelTDX}, {\textsc{amd}}'s
{\textsc{sev}}~\cite{amdSEV21}, and research such as
Sancus~\cite{noorman2013sancus,Noorman:SancusL2017}, 
Sanctum~\cite{CostanLD16}, and Keystone~\cite{Keystone20} for
\textsc{risc-v}.  Schneider et al.'s recent survey on {\tee}s and
their implementation choices~\cite{schneider2022sok} identifies four
main security properties for {\tee}s (p.~1).  One is a weak,
launch-time version of assured remote execution; the second covers our
address space requirements ({\S\,}\ref{sec:caif:services},
\ref{item:if:svc:exec}--\ref{item:if:svc:heap}); the third, concerning
trusted IO, lies outside our current goals; and the fourth is a
data-protection goal to which our {\protfor} provides an elegant
solution.

{\caif} inherits many aspects of previous {\tee}s.  The protected code
segment is available in {\textsc{sgx}}, and is featured in Sancus.  An
intrinsic secret used for key derivation is present in {\textsc{sgx}};
a shared secret like $k_s$ is also in Sancus.  Sancus achieves
software-independent assured remote execution of a particular {\tee}
without asymmetric cryptography in the trust mechanism.

However, our {\protfor} data escrow is distinctive.  The computation
of the {\protfor} key
$\kdf(\mbox{\texttt{"pf"}}, \; \IS, \; \src, \;{\dst})$, while
straightforward, appears not to have been considered in any previous
{\tee} design.  The {\textsc{sgx}} \texttt{egetkey} primitive yields
nothing similar.  The data source obtains a value like
$\kdf(\mbox{\texttt{"c"}}, \; \IS, \; \src)$; the data destination,
receives $\kdf(\mbox{\texttt{"c"}}, \; \IS, \; \dst)$.  These
incomparable keys yield no shared secret; so confidential delivery of
data between {\tee}s seems to require asymmetric cryptography.

Allowing confidential delivery between {\tee}s without any dependence
on long-term asymmetric cryptography is a new contribution of {\caif},
enabling us to
%
install new digital signature algorithms securely long after
anchoring.  

\ifnosecimpl{}\else{ \textsc{sgx}'s \texttt{egetkey} cannot satisfy
    our ideal functionality.  Lemma~\ref{lemma:log:behavior} says of
    {IF} that every successful {\icheck} for $P_s$ is preceded by
    $P_s$ executing an {\iattest}.  If $P_s$ receives a key $k$ via
    \texttt{egetkey} and broadcasts it, then the adversary can create
    {\scmac}s that will pass $\ckat$ without $P_s$ executing {\atloc}.
    This pattern distinguishes the \textsc{sgx}-style mechanism from
    {\IF}.

    Nor can we provide a ``logging enclave'' based on
    \texttt{egetkey}, i.e.~an enclave $P_\ell$ providing reliable
    logging functionality for other enclaves $P_s$.  The obstacle is
    that $P_\ell$ must know $P_s$ has taken the {\atloc}-like action
    to request the log record.  If $P_s$ discloses its
    \texttt{egetkey} key, or any key used to authenticate log requests
    to $P_\ell$, then $P_\ell$ will emit the log record when other
    parties forge log requests.  This again distinguishes the
    \textsc{sgx}-style logging enclave from {\IF}.

    In practice, there is often pressure to act on data in place,
    rather than copying it into local memory first.  This has been the
    source of significant TrustZone attacks, exacerbated because the
    TrustZone ``Secure World'' has memory
    privileges~\cite{MunozRRL23}.  Designers of systems based on
    {\caif} will need care to handle these pressures safely.}  \fi

\subsection{Ideal functionality methods}
\label{sec:related:ideal}

\nocite{GGM86,canetti2001universally}
{\S\,}\ref{sec:properties} shows that {\caif}, when implemented with
strong cryptography, is indistinguishable from the ideal functionality
{\IF}.  Hence our protocol modeling also uses a simple, {{\IF}-style}
state-based treatment of the {\caif} instructions.
Corollary~\ref{cor:caif:if:indistinguishable} ensures that no
tractable observer can tell the difference anyway.

This strategy derives ultimately from Goldreich, Goldwasser, and
Micali's work on random functions~\cite{GGM86}, showing a notion of
pseudorandomness to be indistinguishable from random for any tractable
observer.  Canetti's classic on Universal
Composability~\cite{canetti2001universally} confirms its value
vividly; it shows how to implement many functionalities securely and
justifies their use in all higher level protocols.  An enormous
literature ensued; our
\ifnosecimpl{Corollary~\ref{cor:caif:if:indistinguishable}} \else{
    Theorem~\ref{thm:securely:implements} }\fi
being a small instance of this trend.

\subsection{Protocol analysis}
\label{sec:related:prot:analysis}

\nocite{Guttman10,GuttmanRamsdell19}
\nocite{basin2022tamarin}
\nocite{blanchet2016modeling,bhargavan2021tutorial}

Analyzing security protocols has been a major undertaking since the
late 1970s~\cite{NeedhamSchroeder78}; Dolev and Yao suggested
cryptographic messages be regarded as forming a free algebra and using
symbolic techniques~\cite{DolevYao83}.  A variety of formal approaches
follow them, e.g.~\cite{blanchet2018proverif,meier2013tamarin,cpsa23}.
Computational cryptography also suggests methods for protocol
verification~\cite{BellareRogaway93,CanettiKrawczyk01}.  This raises
the question whether symbolic methods are faithful to the
cryptographic specifics, with a number of approaches yielding
affirmative results in some significant
cases~\cite{AbadiRogaway02,micciancio2004soundness,bana2014computationally}.
Our Thm.~\ref{thm:securely:implements} provides some support for the
soundness of our symbolic protocol analysis.

Our protocols read and write state records.  State raises distinct
problems from messages.  These could be addressed in Tamarin, whose
multiset rewriting model is fundamentally
state-based~\cite{meier2013efficient}.  By contrast, {\cpsa} offers
state in a primarily message-based
formalism~\cite{Guttman12,RamsdellDGR14,guttman2015formal}; for its
current treatment of state, see its manual~\cite[Ch.~8]{cpsa23manual}.
Squirrel interestingly expresses state in a computationally sound
way~\cite{BaeldeEtAl2022}.

{\cpsa} has, helpfully, two modes.  As a \emph{model finder} it
computes the set of essentially different, minimal
executions~\cite{Guttman10}.  This guides protocol design, showing
what is achieved by protocols before they meet their goals.  {\cpsa}
is also a \emph{theorem prover} for security goals proposed by its
user; it produces counterexamples otherwise~\cite{RoweEtAl2016}.


\section{Conclusion}
\label{sec:conc}

{\caif}'s minimal hardware ensures the local service-to-service
\emph{provenance} of data, and its \emph{protection} for known peer
services.  Equipped with strong symmetric cryptoprimitives, {\caif}
provides a secure implementation of an ideal functionality achieving
provenance and protection directly.

We designed a sequence of protocols to run atop {\caif}.  They start
with an initial secure anchoring, a ceremony in a protected space, to
establish a secret $k_s$ shared with an authority.  This key and its
derivatives, protected by {\protfor}, yield channels to known services
on the device.  These channels may be used from distant locations,
e.g.~if the device is on a satellite, to assure remote execution for
new programs.

A delegation service using new algorithms can yield trustworthy
certificate chains for signing keys usable only by known services on
the device.  So assured remote execution can outlast the safety of any
one asymmetric algorithm.

\ifnosecimpl{}\else{
    This is one core need for secure reprogramming:  it
    \emph{authorizes} new programs for remote interactions.  A second
    need is a way to \emph{deauthorize} old programs, blocking
    rollback attacks, in which the adversary benefits by interacting
    with a deprecated version.  Thus, secure reprogramming also needs
    irreversible changes, either to prevent the old programs from
    running, or at least to block access to the keys that previously
    authenticated them.  Although this appears to require little more
    than monotonic counters and constraints on which services can
    advance them, it remains as future work.}
\fi

\nocite{RoweEtAl2016}

{\small
\bibliographystyle{plain}
\bibliography{secureprotocols.bib}

\newpage
\ifllncs{}
\else{
    {\tableofcontents
    \label{toc:page}}
    \newpage}
\fi 
}

\end{document}